\documentclass[traditabstract]{aa}
\usepackage{txfonts}
\usepackage{graphicx}
\usepackage{natbib}
\bibpunct{(}{)}{;}{a}{}{,} % to follow the A&A style
\usepackage{hyperref}
\usepackage{supertabular}
\usepackage{lscape}
\usepackage{subfig}
\usepackage{color}
\def\lsim{\mathrel{\rlap{\lower 3pt\hbox{$\sim$}}\raise 2.0pt\hbox{$<$}}}
\def\gsim{\mathrel{\rlap{\lower 3pt\hbox{$\sim$}} \raise
2.0pt\hbox{$>$}}}

\begin{document}

\title{{\it Herschel}\thanks{{\it Herschel} is an ESA space observatory with science instruments provided by European-led Principal Investigator consortia and with important participation from NASA.}-ATLAS: {\it Planck} sources in the Phase 1 fields
%  \thanks{}
\subtitle{} }
   \author{D. Herranz\inst{1,2}\fnmsep\thanks{e-mail: herranz@ifca.unican.es}
          \and J. Gonz\'alez-Nuevo\inst{1,3} 
  	      \and D.L. Clements\inst{4}
          \and M. Clemens\inst{5}
          \and G. De Zotti\inst{5,3}
          \and M. Lopez-Caniego\inst{1}
          \and A. Lapi\inst{6,3}
          \and G. Rodighiero\inst{7}
          \and L. Danese\inst{3}
          \and H. Fu\inst{8}
          \and A. Cooray\inst{8}
          \and M. Baes\inst{9}
          \and G.~J. Bendo\inst{10}
          \and L. Bonavera\inst{1,3}    
       	  \and F.~J. Carrera\inst{1}
          \and H. Dole\inst{11} 
          \and S. Eales\inst{12}
          \and R.\ J.\ Ivison\inst{13,14}  
          \and M. Jarvis\inst{15,16}
          \and G. Lagache\inst{11}
          \and M. Massardi\inst{17}
          \and M.\ J.\ Micha{\l}owski\inst{14}
          \and M. Negrello\inst{5}
          \and E. Rigby\inst{18}
          \and D. Scott\inst{19}
          \and E. Valiante\inst{12}
          \and I. Valtchanov\inst{20}
          \and P. Van der Werf\inst{18}  
          \and R. Auld\inst{12}
          \and S. Buttiglione\inst{5}
          \and A. Dariush\inst{4}
          \and L. Dunne\inst{21}
          \and R. Hopwood\inst{4}
          \and C. Hoyos\inst{21}
          \and E. Ibar\inst{13}
          \and S. Maddox\inst{21}
}	
\institute{Instituto de F\'\i{sica} de Cantabria (CSIC-UC), Avda. los Castros s/n, 39005 Santander, Spain
       \and Astrophysics Group, Cavendish Laboratory, University of Cambridge, Cambridge CB3 0HE
       \and Astrophysics Sector, SISSA, Via Bonomea 265, 34136 Trieste, Italy
       \and Astrophysics Group, Imperial College, Blackett Laboratory, Prince Consort Road, London SW7 2AZ, UK
	   \and INAF-Osservatorio Astronomico di Padova, Vicolo dell'Osservatorio 5, I-35122 Padova, Italy
       \and Dipartimento di Fisica, Universit\`a di Roma `Tor Vergata', Via Ricerca Scientifica 1, 00133 Roma, Italy
       \and Dip. Astronomia, Univ. di Padova, Vicolo dell'Osservatorio 3, I-35122 Padova, Italy
       \and Department of Physics and Astronomy, Frederick Reines Hall, University of California, Irvine, CA 92697--4575, USA
       \and Sterrenkundig Observatorium, Universiteit Gent, Krijgslaan 281 S9, B-9000 Gent, Belgium
       \and UK ALMA Regional Centre Node, Jodrell Bank Centre for Astrophysics, School of Physics and Astronomy, University
            of Manchester, Oxford Road, Manchester M13 9PL, United Kingdom 
       \and Institut dAstrophysique Spatiale (IAS), b\^atiment 121, Universit\'e Paris-Sud 11 and CNRS (UMR 8617), 91405 Orsay, France
   	   \and Cardiff School of Physics and Astronomy, Cardiff University, Queens Building, The Parade, Cardiff, CF24 3AA, UK
   	   \and UK Astronomy Technology Centre, Royal Observatory, Blackford Hill, Edinburgh EH9 3HJ
   	   \and SUPA\thanks{Scottish Universities Physics Alliance}, Institute for Astronomy, University of Edinburgh, Royal Observatory, 
   	        Edinburgh, EH9 3HJ, UK
   	   \and Centre for Astrophysics Research, Science \& Technology Research Institute, University of Hertfordshire, Hatfield, Herts, AL10 9AB, UK
	   \and Physics Department, University of the Western Cape, Cape Town, 7535, South Africa	
       \and INAF-Istituto di Radioastronomia, via Gobetti 101, 40129 Bologna, Italy
       \and Leiden Observatory, Leiden University, P.O. Box 9513, 2300 RA Leiden, The Netherlands
       \and Department of Physics \& Astronomy, University of British Columbia, 6224 Agricultural Road, Vancouver, British Columbia, Canada
       \and Herschel Science Centre, ESAC, ESA, P.O. Box 78, Villanueva de la Ca\~nada, 28691 Madrid, Spain
	   \and School of Physics and Astronomy, University of Nottingham, University Park, Nottingham NG7 2RD, UK
%		\and Sterrenkundig Observatorium, Universiteit Gent, Krijgslaan 281 S9, B-9000 Gent, Belgium   
%   \and Institute for Astronomy, University of Edinburgh, Royal Observatory, Blackford Hill, Edinburgh EH9 3HJ
%   \and Centre for Astrophysics Research, Science \& Technology Research Institute, University of Hertfordshire, Hatfield, Herts, AL10 9AB, UK
%   \and Astrophysics Branch, NASA/Ames Research Center, MS 245-6, Moffett Field, CA 94035, USA
           }

%\offprints{}

\date{}

\abstract{We present the results of a cross-correlation of the \emph{Planck} Early Release Compact Source Catalog (ERCSC) with the catalog of \emph{Herschel}-ATLAS sources detected in the Phase 1 fields, covering $134.55\,\hbox{deg}^2$. There are 28 ERCSC sources detected by \emph{Planck}  at 857 GHz in this area. As many as 16 of them are probably high Galactic latitude cirrus; 10 additional sources can be clearly identified as bright, low-$z$ galaxies; one further source is resolved by \emph{Herschel} as two relatively bright sources; and the last is resolved into an unusual condensation of low-flux, probably high-redshift point sources, around a strongly lensed \emph{Herschel}-ATLAS source at $z=3.26$.  Our results demonstrate that the higher sensitivity and higher angular resolution H-ATLAS maps provide essential  information for the interpretation of candidate sources extracted from \emph{Planck} sub-mm maps.}


\maketitle

\section{Introduction} \label{sec:intro}

During the past year, the simultaneous operation of ESA's \emph{Herschel} \citep{HERSCHEL} and \emph{Planck} \citep{Planck,Planck1} missions has given us an unprecedented opportunity to cover one of the last few observational gaps in the far-infrared and sub-millimeter regions of the electromagnetic spectrum. \emph{Herschel} is an observatory facility that covers the $55 - 671$ $\mu$m spectral range, with angular resolution ranging between 6 and 35 arcseconds \citep{HERSCHEL}. \emph{Planck} is a surveyor that is observing the whole sky in nine spectral bands between 350 $\mu$m and 1 cm and with angular resolution ranging from $4.23$ to $32.65$ arcmin. \emph{Planck} has two frequency channels close to \emph{Herschel} bands: the 545 and 857 GHz (550 and $350\,\mu$m) channels of the High Frequency Instrument (HFI). In this paper we will study the cross-correlation of the \emph{Planck} Early Release Compact Source Catalog (ERCSC) with the Phase 1 of the catalogue of the \emph{Herschel} Astrophysical Terahertz Large Area Survey \citep[H-ATLAS,][]{HATLAS}.

The overlap in time and frequency between \emph{Herschel} and \emph{Planck} is not accidental: the two missions have been planned while keeping in mind the added scientific value of a synergy between them \citep{bluebook}. In addition to providing a broader spectral coverage of common sources, the combination of  \emph{Planck} and \emph{Herschel} data will be beneficial in other respects. In particular, the much higher resolution and sensitivity of  \emph{Herschel} makes it well suited for follow-up of \emph{Planck} sources\footnote{For comparison, the resolution at the SPIRE 350 $\mu$m band is FWHM=29.4 arcsec, whereas the ERCSC nominal beam at 857 GHz is 4.23 arcmin.}, allowing us to assess the effects of source confusion in \emph{Planck} channels. 
In some cases it will be possible to resolve individual detections by \emph{Planck} into separated sources. More generally, \emph{Herschel} will make it possible to quantify the boosting of \emph{Planck} fluxes by the many faint sources within its beam.  Moreover, \emph{Herschel} data can be used to improve the foreground characterization, thus helping to distinguish between genuine, possibly extragalactic, point  sources and compact Galactic cirrus, and to provide more precise positions, essential for source identification at other wavelengths. This knowledge will be important for the interpretation of the all-sky \emph{Planck} survey data. Note however that although \textit{Herschel}'s resolution is much better than \textit{Planck}'s, it is still highly likely that many of the \textit{Herschel} 350 and 500 $\mu$m sources are also blends. H-ATLAS maps are in general not affected by source confusion, except in regions with important cirrus, but \textit{Herschel} sources resolve to multiple MIPS \citep[Multiband
Imaging Photometer for \textit{Spitzer}][]{rieke04} sources in many cases.  

In turn, \emph{Herschel} will benefit from the absolute calibration of \emph{Planck} fluxes which should be better than 2\% up to 353 GHz, where it is based on the CMB dipole, and $\simeq 7\%$ in the two highest frequency channels (545 and 857 GHz), where it is based on a comparison with COBE/FIRAS \citep{HFI_dataproc,LFI_dataproc}\footnote{This refers to the absolute calibration of the instrument. For individual sources, however, the uncertainty in the photometry is affected by other factors such as the uncertainty on the beam shape and the possibly extended nature of the source, leading to errors that can be as large as 30\%, as cited by the \emph{Explanatory Supplement to the Planck ERCSC}.}.  For comparison, the overall photometric accuracy of the \emph{Herschel}-SPIRE instrument is conservatively estimated as $\simeq 7\%$ \citep{spire_manual}\footnote{The SPIRE Observers\textquoteright \ Manual is available from the \emph{Herschel}
Science Centre: 
\newline
 \url{http://herschel.esac.esa.int/Docs/SPIRE/pdf/spire_om.pdf}.}.
 A comparison between catalogues of galaxies observed with \emph{Herschel} and \emph{Planck} can be used as a check on the calibration of the two observatories, which are done in different ways.

\begin{figure*}
  \resizebox{\hsize}{!}{\includegraphics[angle=270]{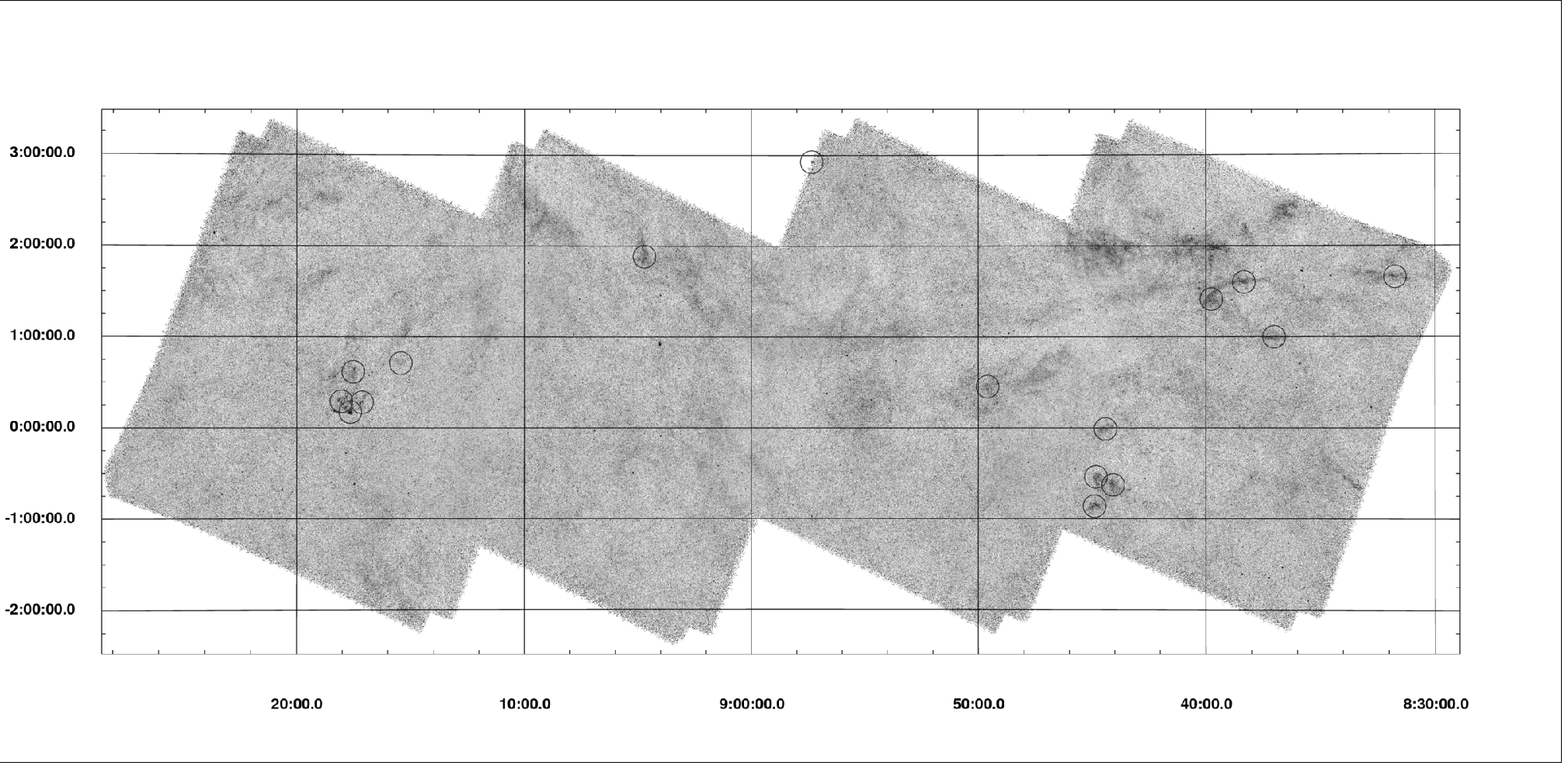}}
  \caption{Positions of the ERCSC 857 GHz sources in the 350 $\mu$m ATLAS GAMA09 field. }
  \label{fig:GAMA-09}
\end{figure*}

\begin{figure*}
  \resizebox{\hsize}{!}{\includegraphics[angle=270]{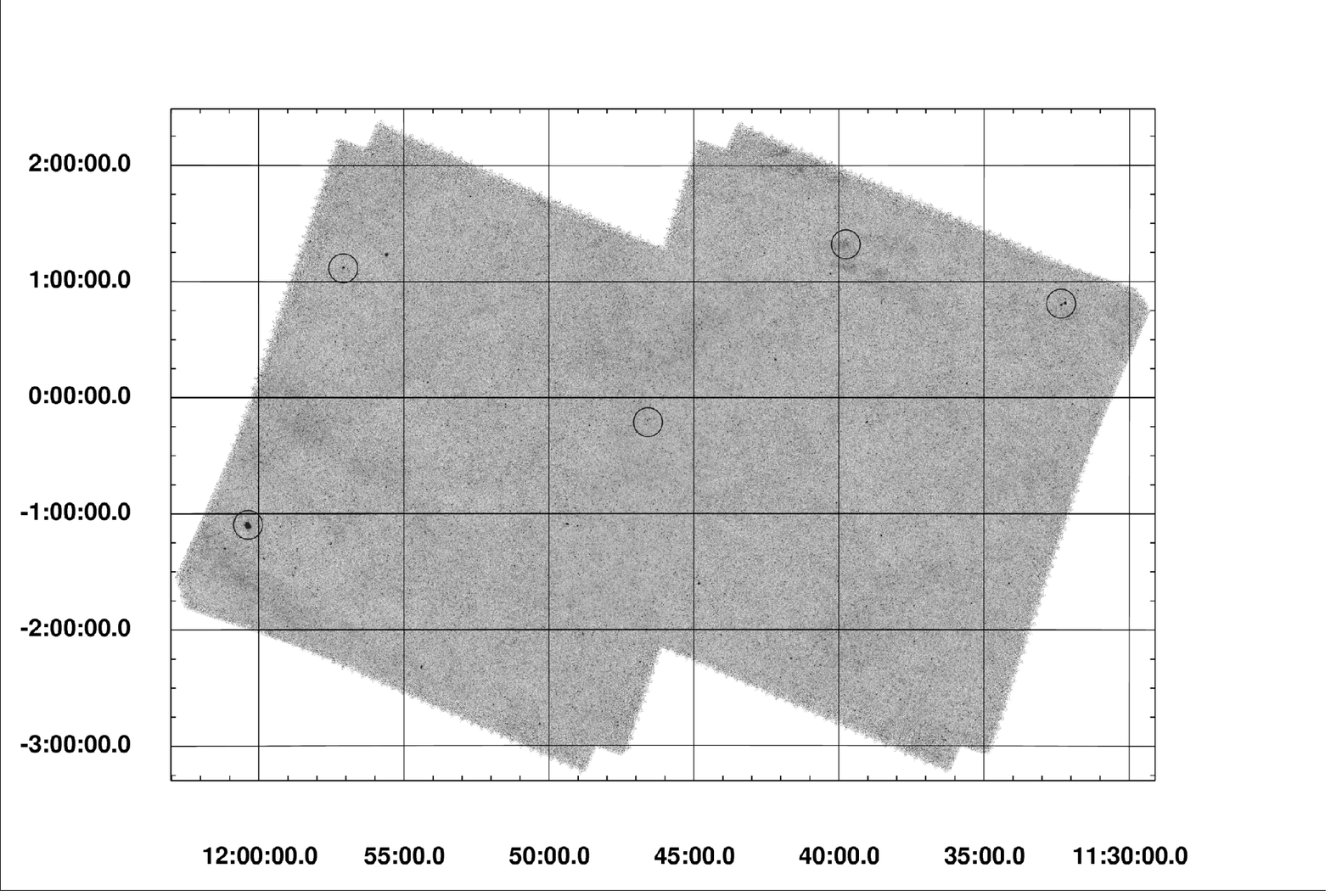}}
  \caption{Positions of the ERCSC 857 GHz sources in the 350 $\mu$m ATLAS GAMA-12 field. }
  \label{fig:GAMA-12}
\end{figure*}

\begin{figure*}
  \resizebox{\hsize}{!}{\includegraphics[angle=270]{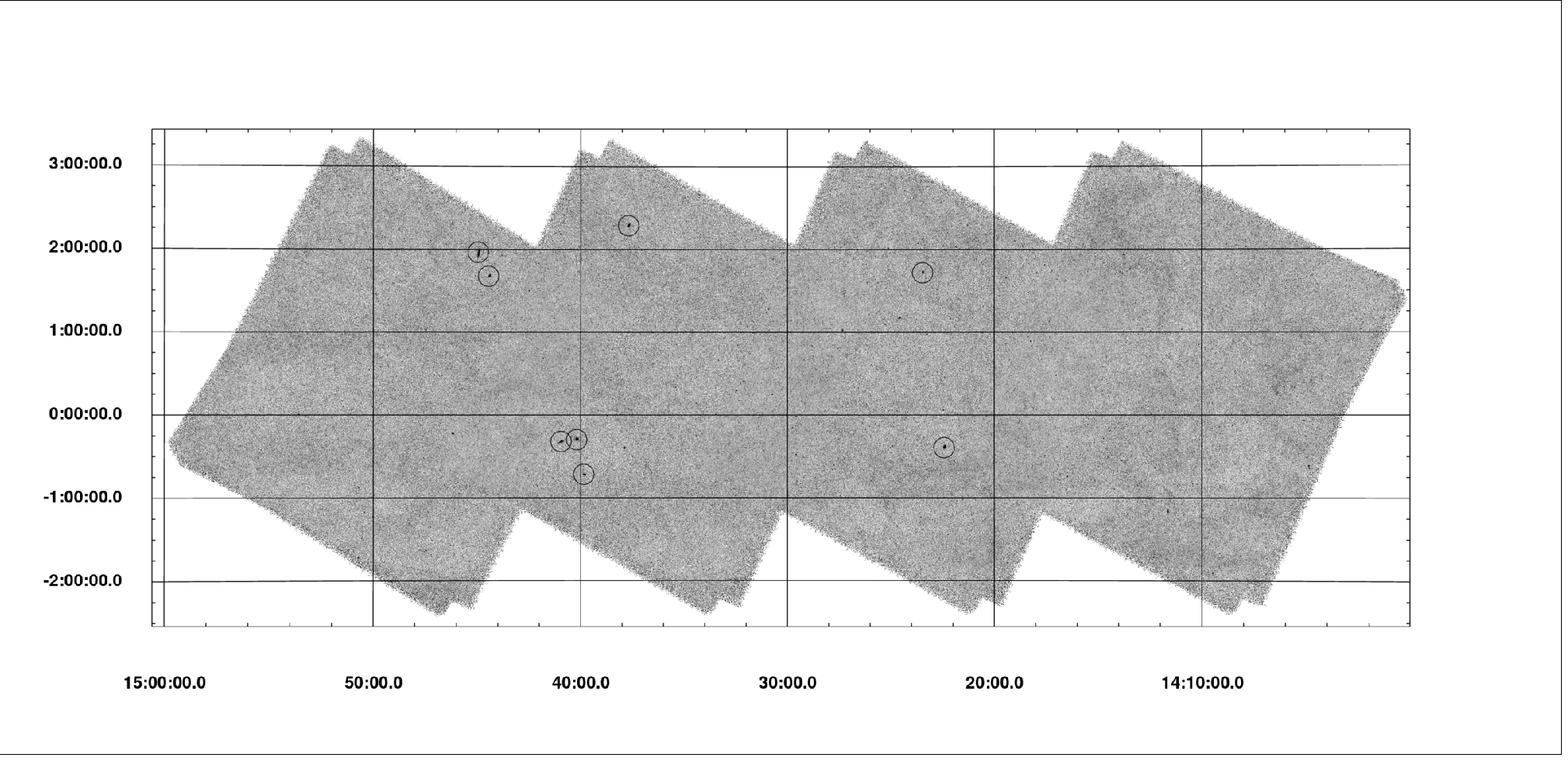}}
  \caption{Positions of the ERCSC 857 GHz sources in the 350 $\mu$m ATLAS GAMA-15 field. }
  \label{fig:GAMA-15}
\end{figure*}

\begin{figure*}
  \centering
  \resizebox{0.8\hsize}{!}{\includegraphics[angle=0]{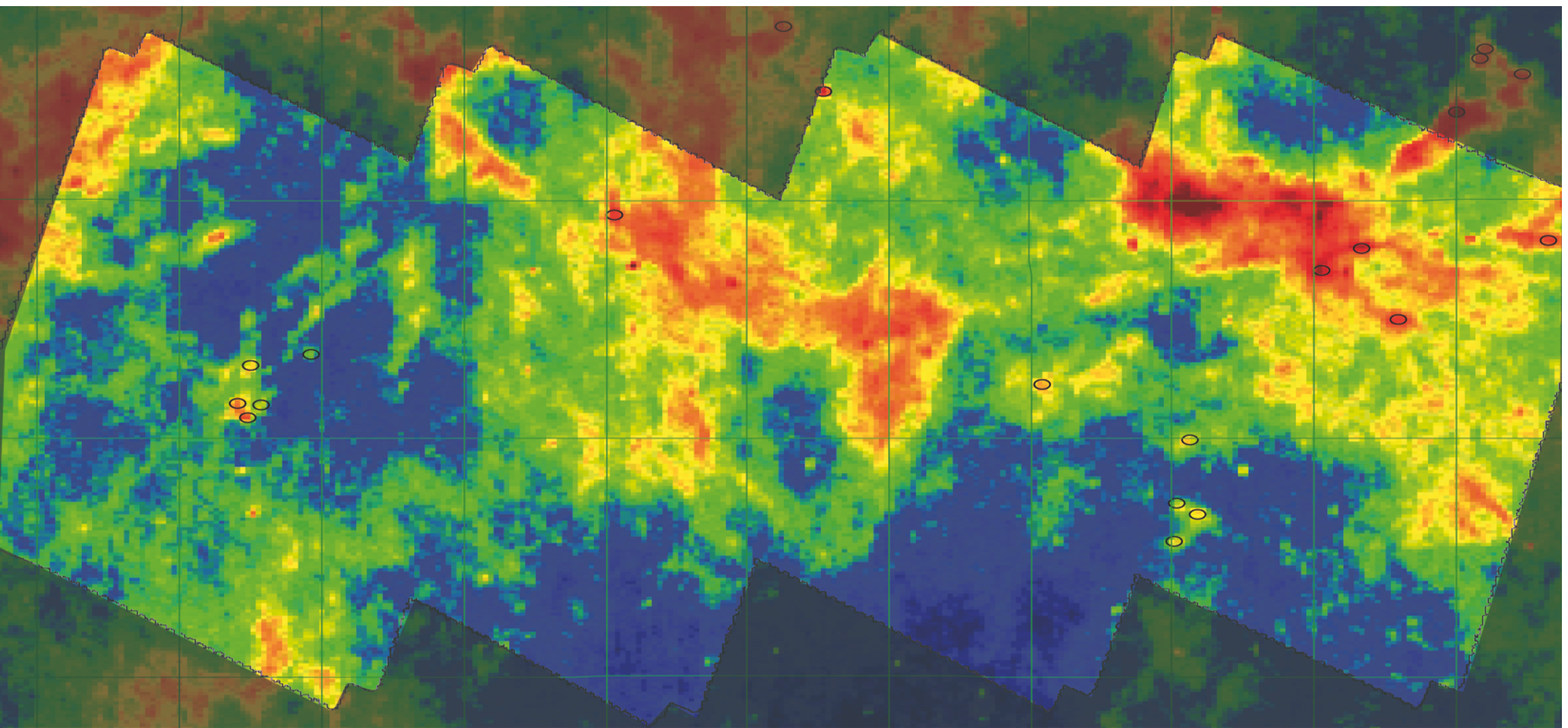}}
  \caption{Position of the ERCSC sources (black ovals) superimposed on the IRIS 100 micron map around the GAMA-09 H-ATLAS field. }
  \label{fig:iris}
\end{figure*}

Due to their different angular resolution and sensitivity, \emph{Planck} and \emph{Herschel} have almost complementary selection functions: \emph{Planck} is most sensitive to either local galaxies or extreme high redshift objects whereas most H-ATLAS sources lie around $z \sim 1$ \citep{HATLAS}. Moreover, \emph{Planck} covers a broader spectral range (nine frequencies from 30 to 857 GHz), which is useful in order to improve the characterization and removal of foregrounds and to follow the SEDs of interesting objects, such as blazars, down to radio frequencies.
With its all-sky coverage, \emph{Planck} is ideal for detecting the rarest, most extreme (sub-)mm sources. In particular, it may detect the most luminous proto-clusters of dusty galaxies, whose  luminosities, integrated over the \emph{Planck}  beam, may be, at $z \gtrsim 1$, more than an order of magnitude higher than the mean luminosity of individual dusty galaxies at the same redshift \citep{negrello05}. The far superior \emph{Herschel} resolution and
point source detection capabilities will then allow us to establish the nature of candidate high-$z$ proto-clusters and to characterize the physical properties of those that are confirmed.

This latter point deserves some more attention. The discovery of distant far-IR luminous galaxies by sub-mm imagers \citep[e.g.][]{smail97,hughes98} and the discovery of the Cosmic Infrared Background \citep{puget96,fixsen98} have demonstrated the importance of the far-IR/sub-mm bands in determining a complete picture of the history of galaxy formation and evolution. The high redshift ($z\sim$2--3) sources detected in these sub-mm surveys are expected to be the progenitors of the giant elliptical galaxies that we see today \citep[e.g.][]{Blain2002}. In the framework of  hierarchical clustering models of large scale structure and galaxy formation we would expect that the most massive elliptical galaxies form in the cores of what will become today's richest galaxy clusters.
 \cite{granato04} and others predict that multiple galaxies in such regions will undergo dust obscured starbursts at about the same time, producing clumps of dusty proto-spheroidal galaxies.  Moreover, 
the ages of elliptical galaxies in low-$z$ clusters are all similar, which implies that they all probably formed in large and rapid starburst events at high-$z$ \citep{propis99}, probably associated with clumps of forming galaxies.

Hints of such objects may already have been found by clustering studies with Spitzer \citep[e.g.][]{magl07}, or using high-$z$ quasars as signposts for possible proto-clusters \citep{ivison2000,stevens03,priddey08,stevens10}. The latter study finds far more sub-mm bright companions to quasars than expected from the blank-field number counts, suggesting the presence of dusty proto-clusters. Meanwhile, the highest redshift proto-cluster currently known, at $z\sim 5.3$, includes at least one sub-mm luminous object \citep{capak11}. While this object is extreme, such sources may need to be quite common if the recent discovery of a mature galaxy cluster at $z=2.07$ \citep{gobat11}, with a fully formed red-sequence of galaxies with ages  $> 1.3$ Gyr, is representative of a significant population.

The dusty star-forming phase of a proto-cluster is likely to be quite short, so the objects should be rare on the sky. Fortunately, \emph{Herschel} and \emph{Planck} are up to this challenge. \emph{Herschel} allows relatively large areas of the sky to be covered to sensitive flux levels at wavelengths corresponding to the peak of the dust spectral energy distribution (SED) of high redshift starbursts \citep[e.g.][]{Lapi2011}. Meanwhile, \emph{Planck} provides all-sky coverage at wavelengths matching the longest \textit{Herschel} bands and stretching
into the mm. Therefore, \textit{Planck} can find cold compact structures and \textit{Herschel} can then confirm that these are associated with clumps of galaxies, potentially at high redshift.

The Early Release Compact Source Catalogue \citep[ERCSC,][]{ERCSC} provides an all-sky list of compact Galactic and extragalactic objects including stars with dust shells, prestellar cores, radio galaxies, blazars, infrared luminous galaxies, Galactic interstellar medium features, cold molecular cloud core candidates and galaxy cluster candidates. The list contains more than 15,000 distinct objects, $\sim 60\%$ of which are visible  in the  \emph{Planck} highest frequency (545 and 857 GHz) channels that virtually overlap with the \emph{Herschel}/SPIRE bands. A sufficiently wide-area survey made with \emph{Herschel} is bound to contain at least some of these sources.

The  \emph{Herschel} Astrophysical Terahertz Large Area Survey {H-ATLAS} \citep{HATLAS} is the largest area survey carried out by the \emph{Herschel} Space Observatory   \citep{HERSCHEL}.  It will cover $\sim 550\,\mathrm{deg}^2$ with PACS \citep{pog10} and SPIRE \citep{griffin10} in five bands, from $100$ to $500\,\mu$m. The Phase 1 observations have surveyed enough area ($134.55\,\mathrm{deg}^2$) to allow a preliminary, yet meaningful comparison with  \emph{Planck} ERCSC data.

In this paper we present the results of a cross-correlation of the \emph{Planck} ERCSC catalog with the catalog of H-ATLAS sources detected in the Phase 1 fields (Dunne et al., in preparation).
The common sources are described in \S\,\ref{sec:sources}.  In \S\,\ref{sec:photo} we compare the flux density estimates of both experiments. In \S\,\ref{sec:cirrus} we study the contamination of the ERCSC subsample by looking for extended diffuse emission as a tracer of high-latitude cirrus. In \S\,\ref{sec:proto} we discuss a very unusual source that may be a combination of a (maybe random) condensation of low redshift low-flux galaxies and a strongly lensed galaxy. Finally, in \S\,\ref{sec:conclusions} we summarize our conclusions.

\section{Planck sources in the H-ATLAS phase 1 fields} \label{sec:sources}

\begin{figure*}

  \centering
  
  \subfloat[G223.40+22.96]{\includegraphics[width=0.25\textwidth]{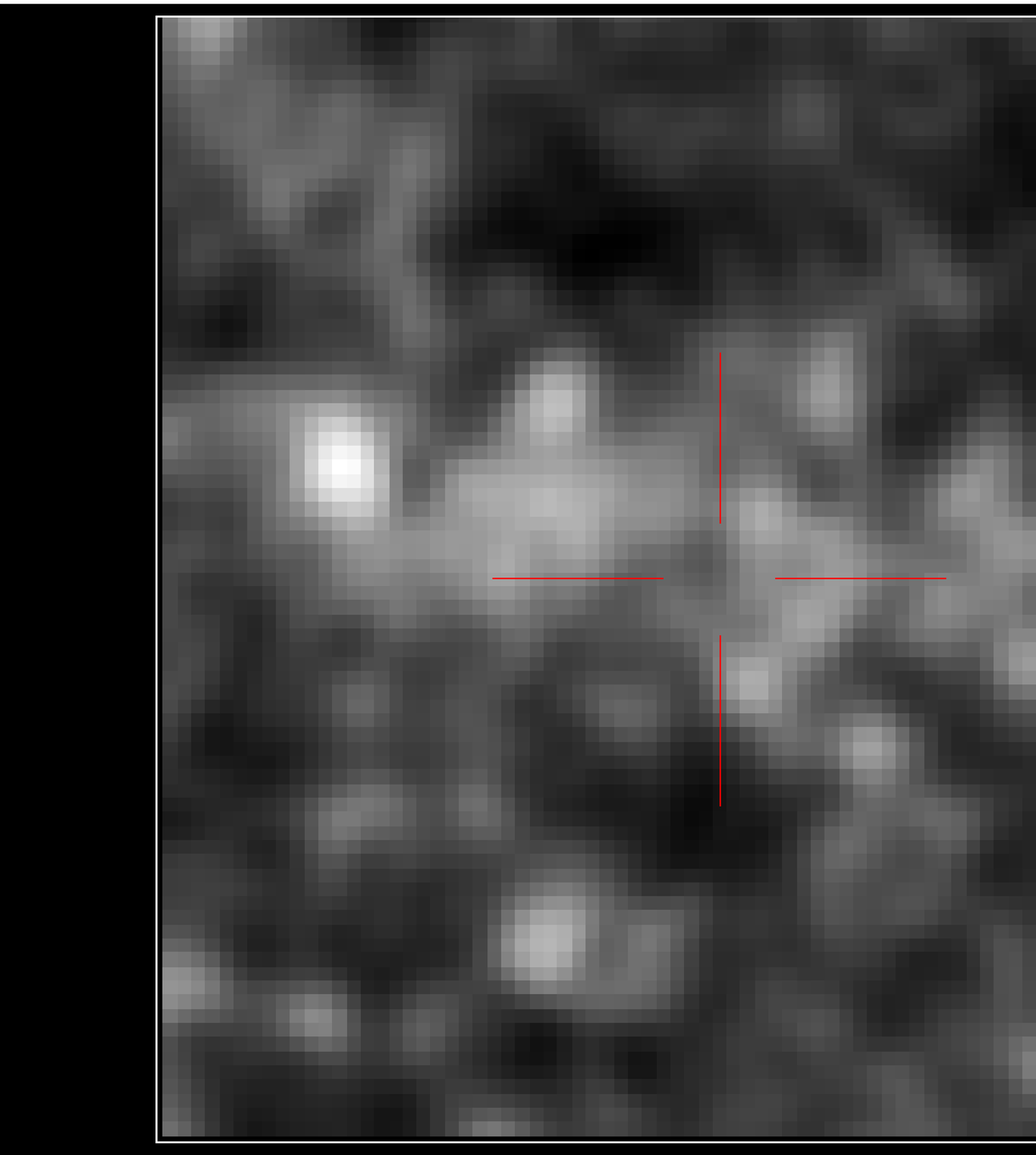}}
  \subfloat[G224.33+24.38]{\includegraphics[width=0.25\textwidth]{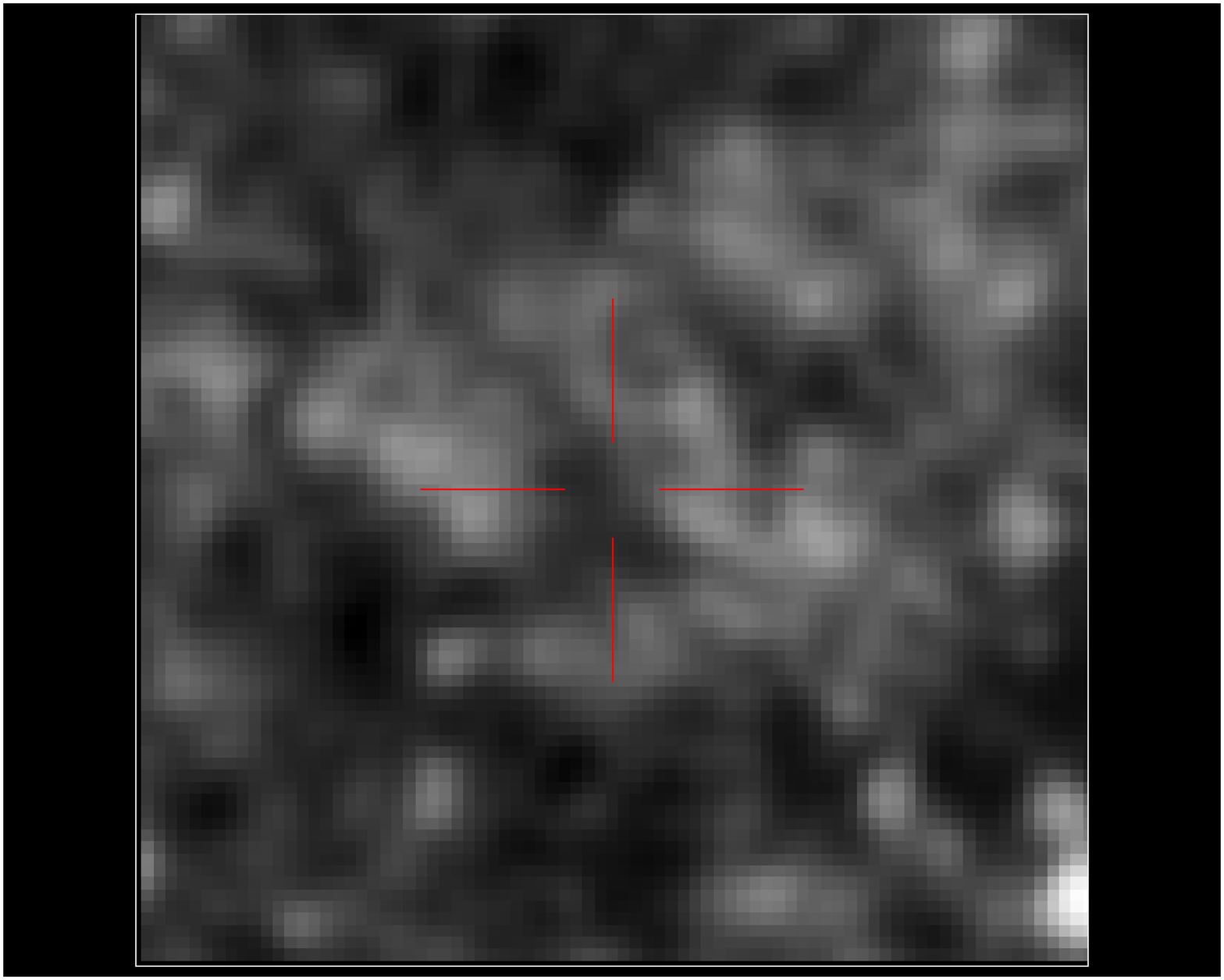}}
  \subfloat[G224.70+24.60]{\includegraphics[width=0.25\textwidth]{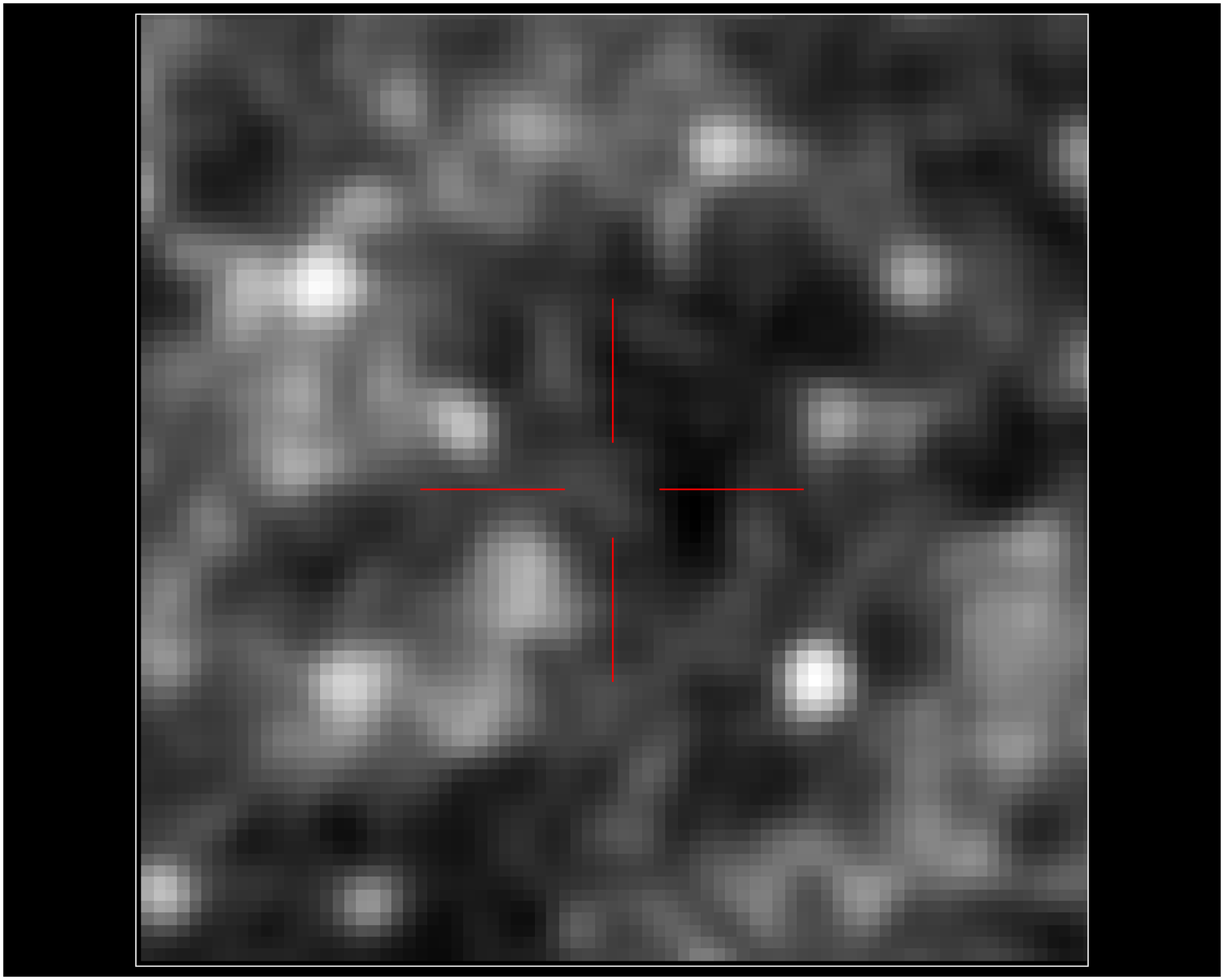}}
  \subfloat[G224.73+23.79]{\includegraphics[width=0.25\textwidth]{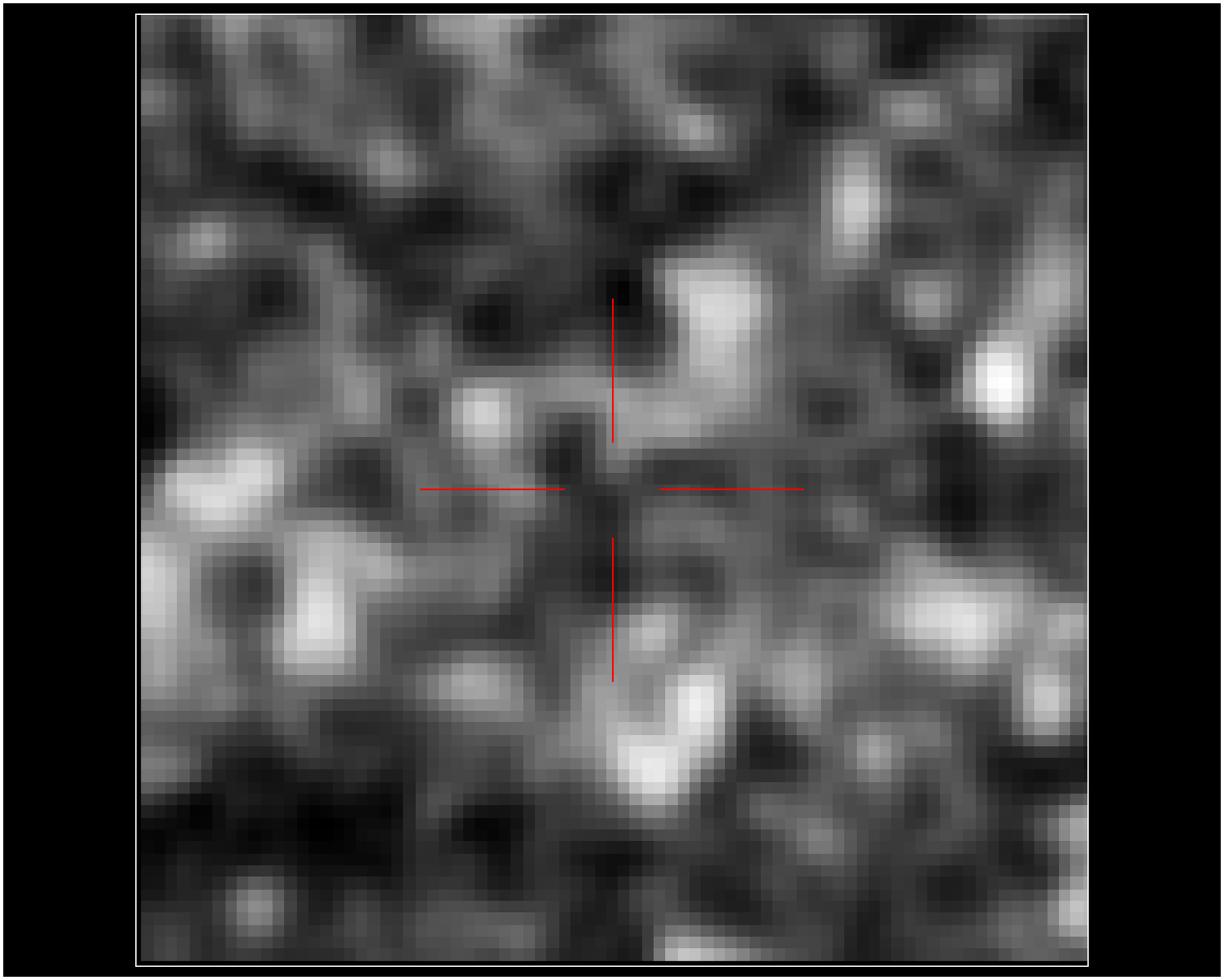}} 
  \\
  
  \subfloat[G226.70+24.89]{\includegraphics[width=0.25\textwidth]{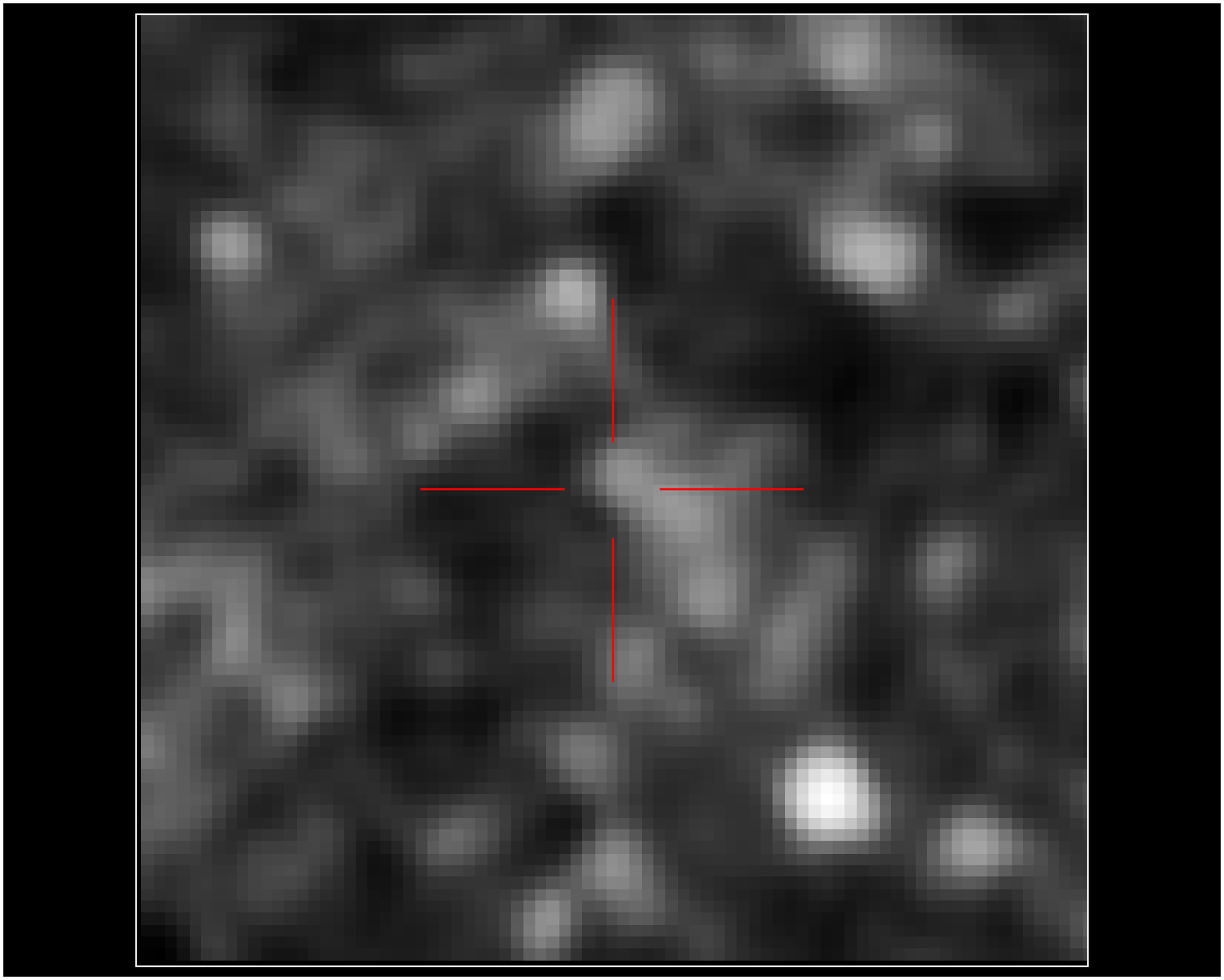}}    
  \subfloat[G226.98+26.25]{\includegraphics[width=0.25\textwidth]{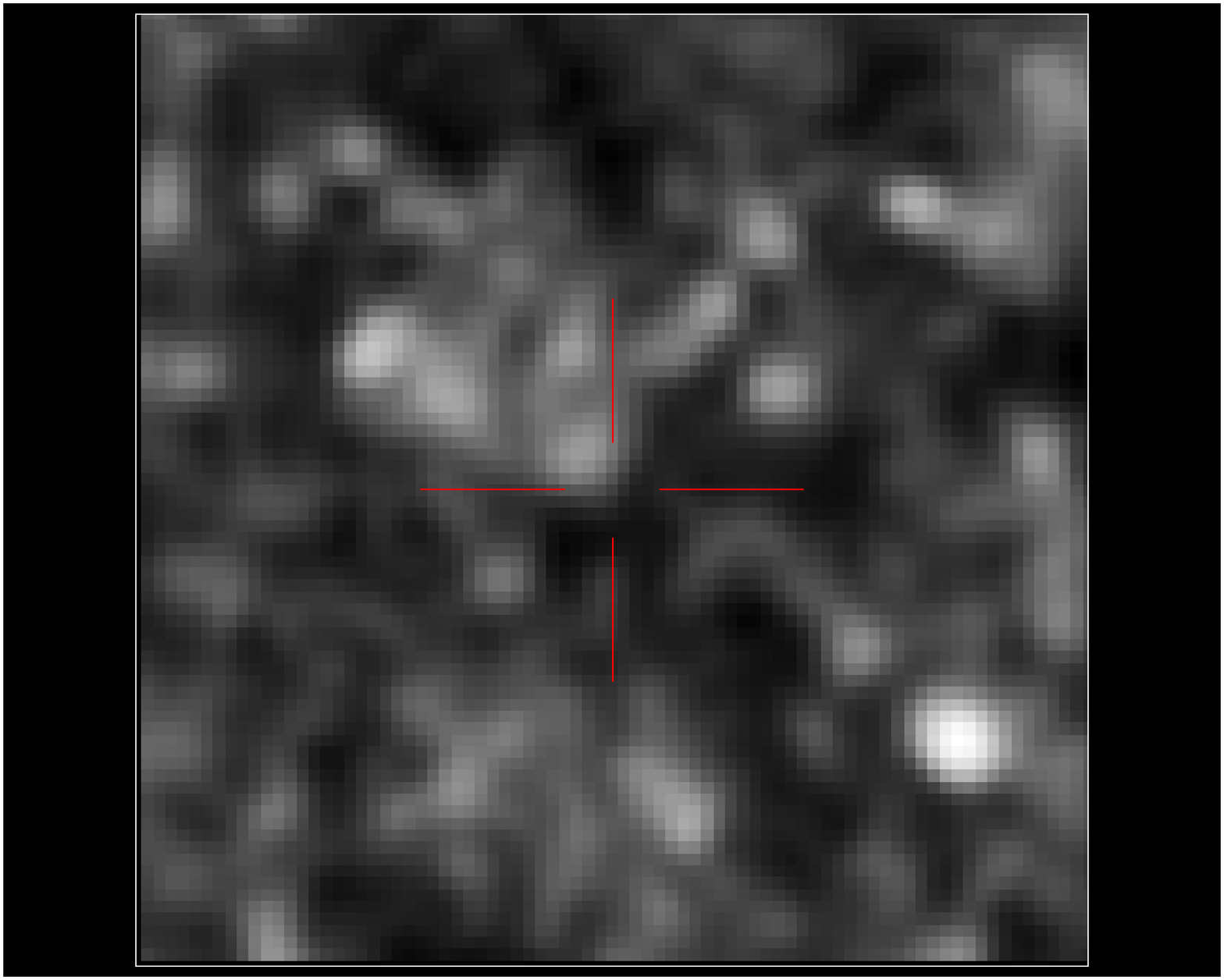}}
  \subfloat[G227.24+24.51]{\includegraphics[width=0.25\textwidth]{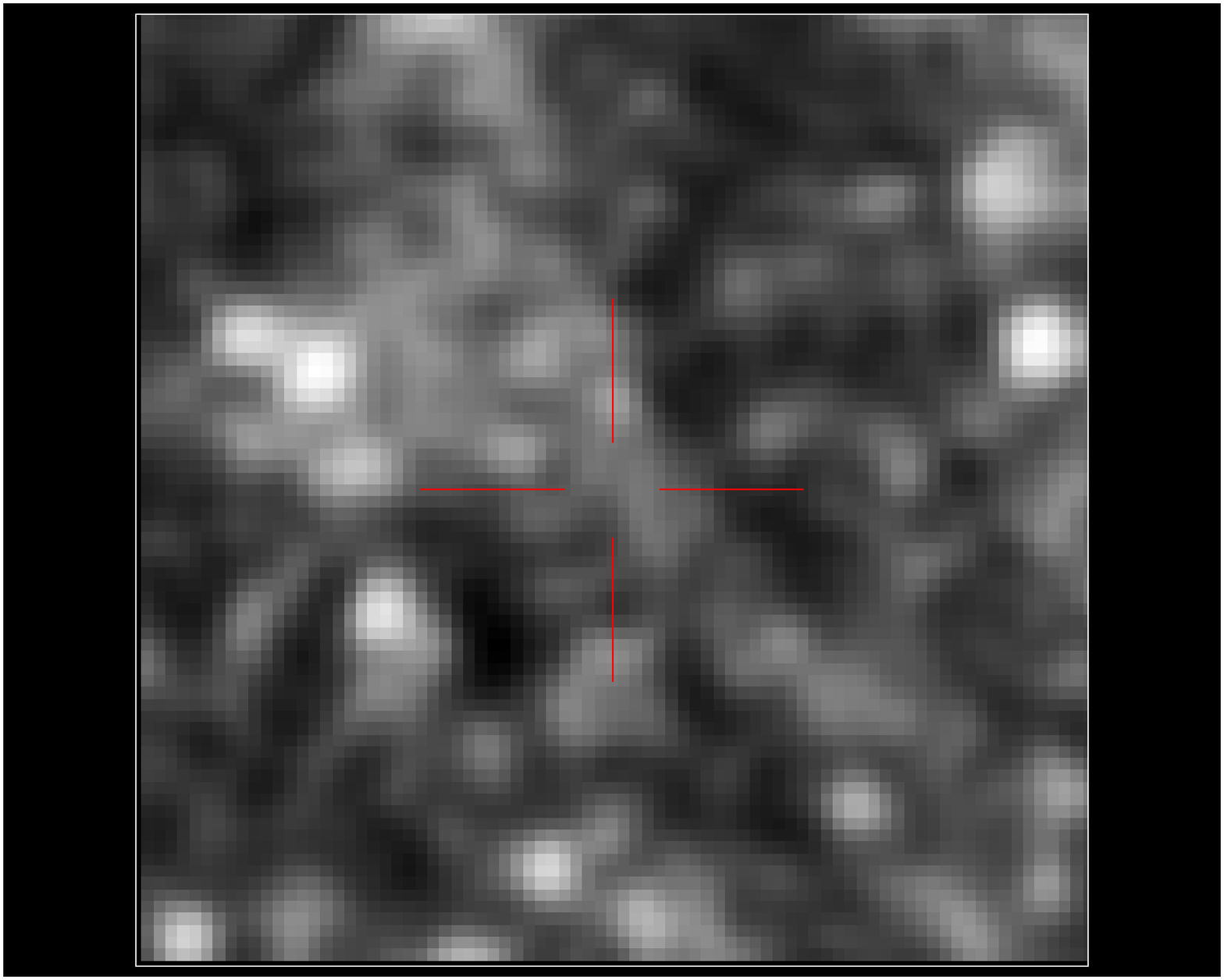}}
  \subfloat[G227.26+24.72]{\includegraphics[width=0.25\textwidth]{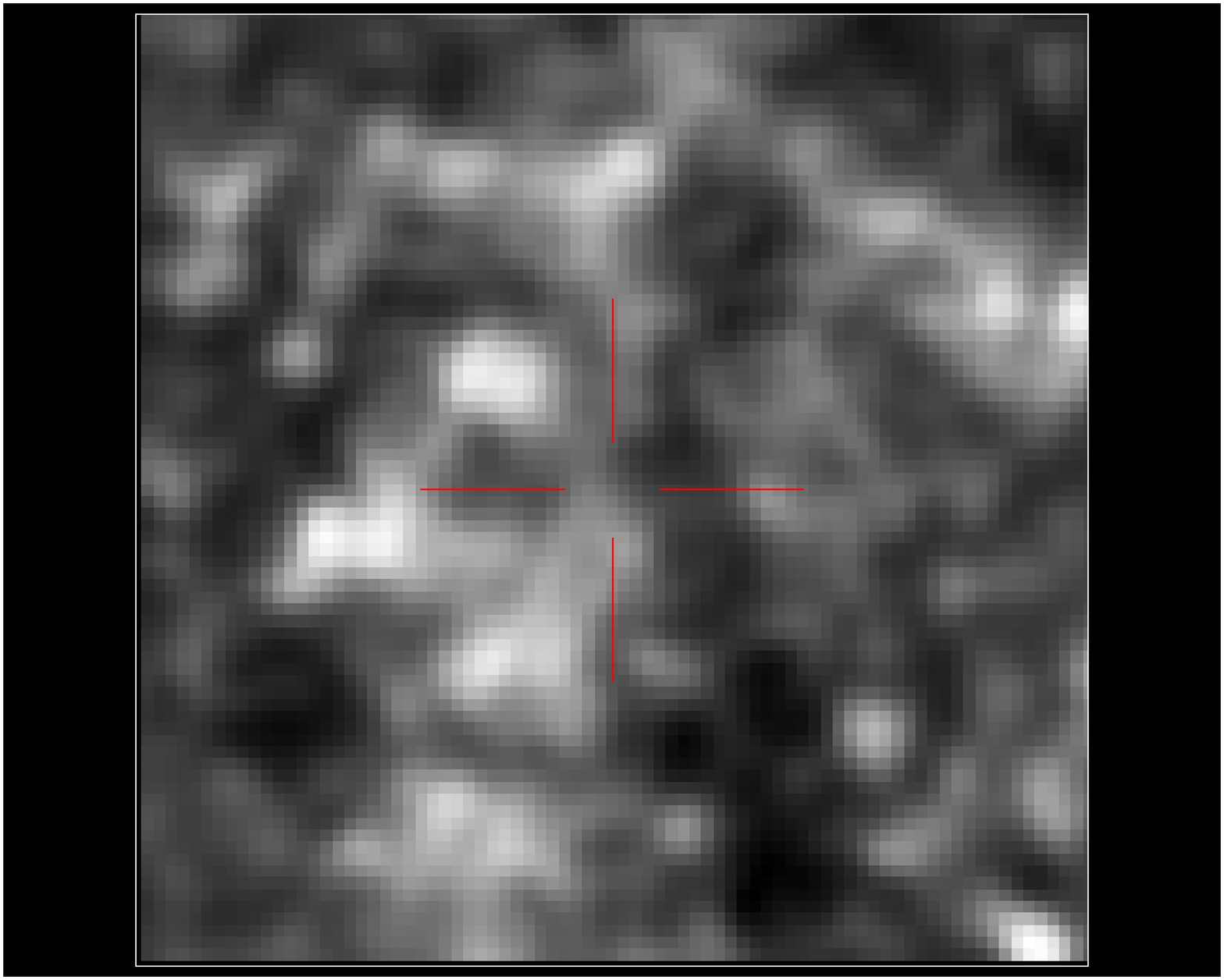}} 
  \\
 
  \subfloat[G227.58+24.57]{\includegraphics[width=0.25\textwidth]{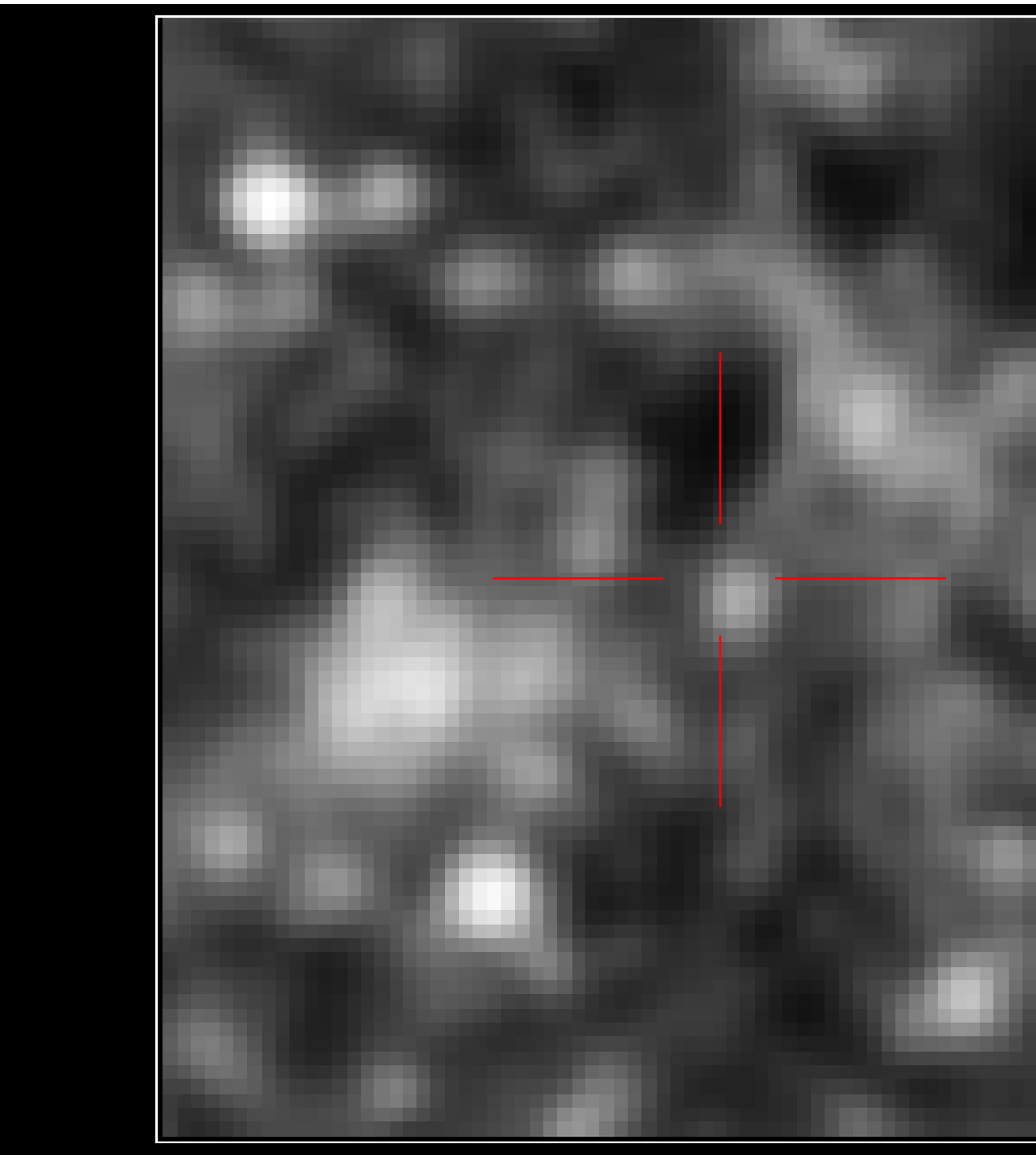}}
  \subfloat[G227.75+30.24]{\includegraphics[width=0.25\textwidth]{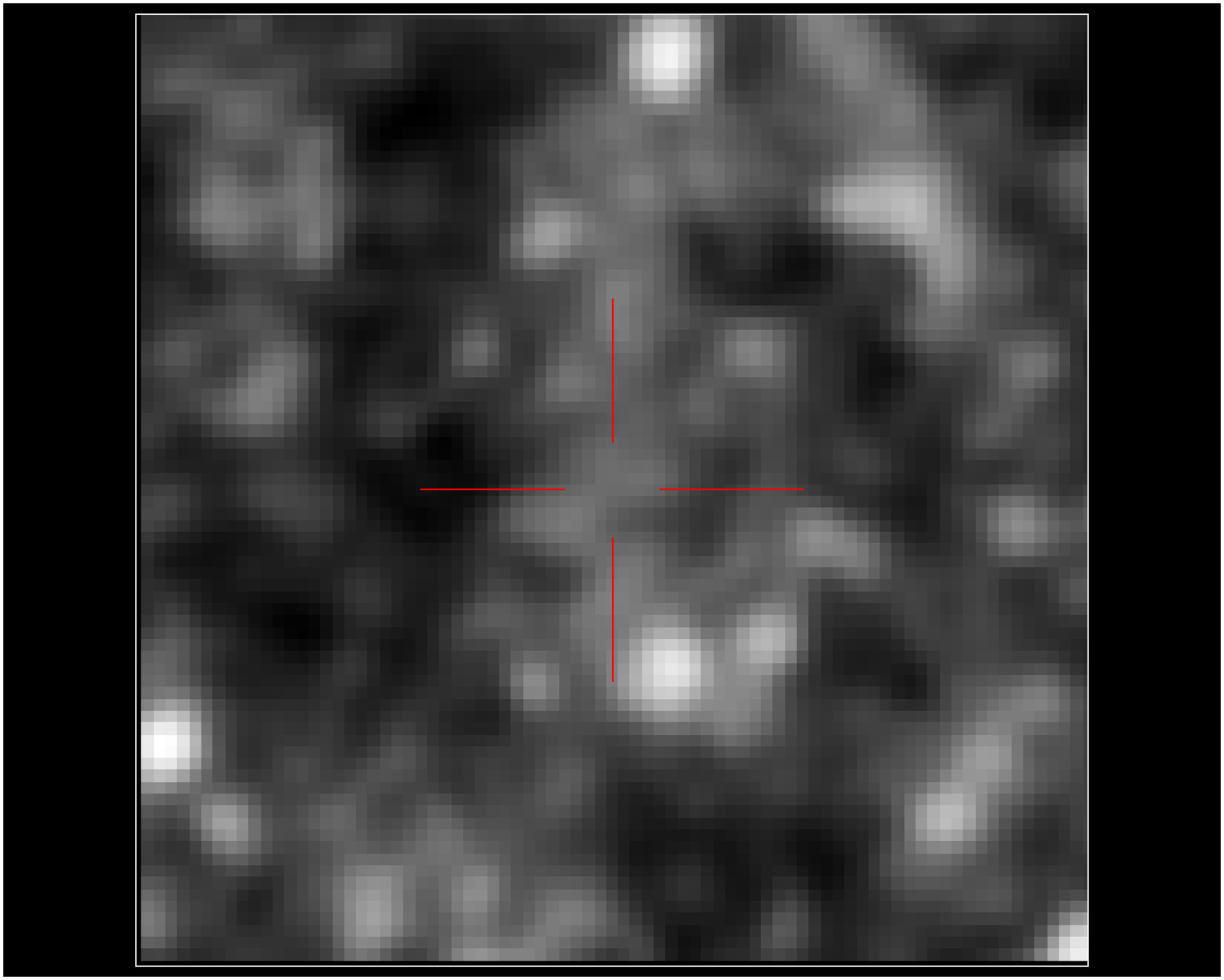}}
  \subfloat[G230.55+31.91]{\includegraphics[width=0.25\textwidth]{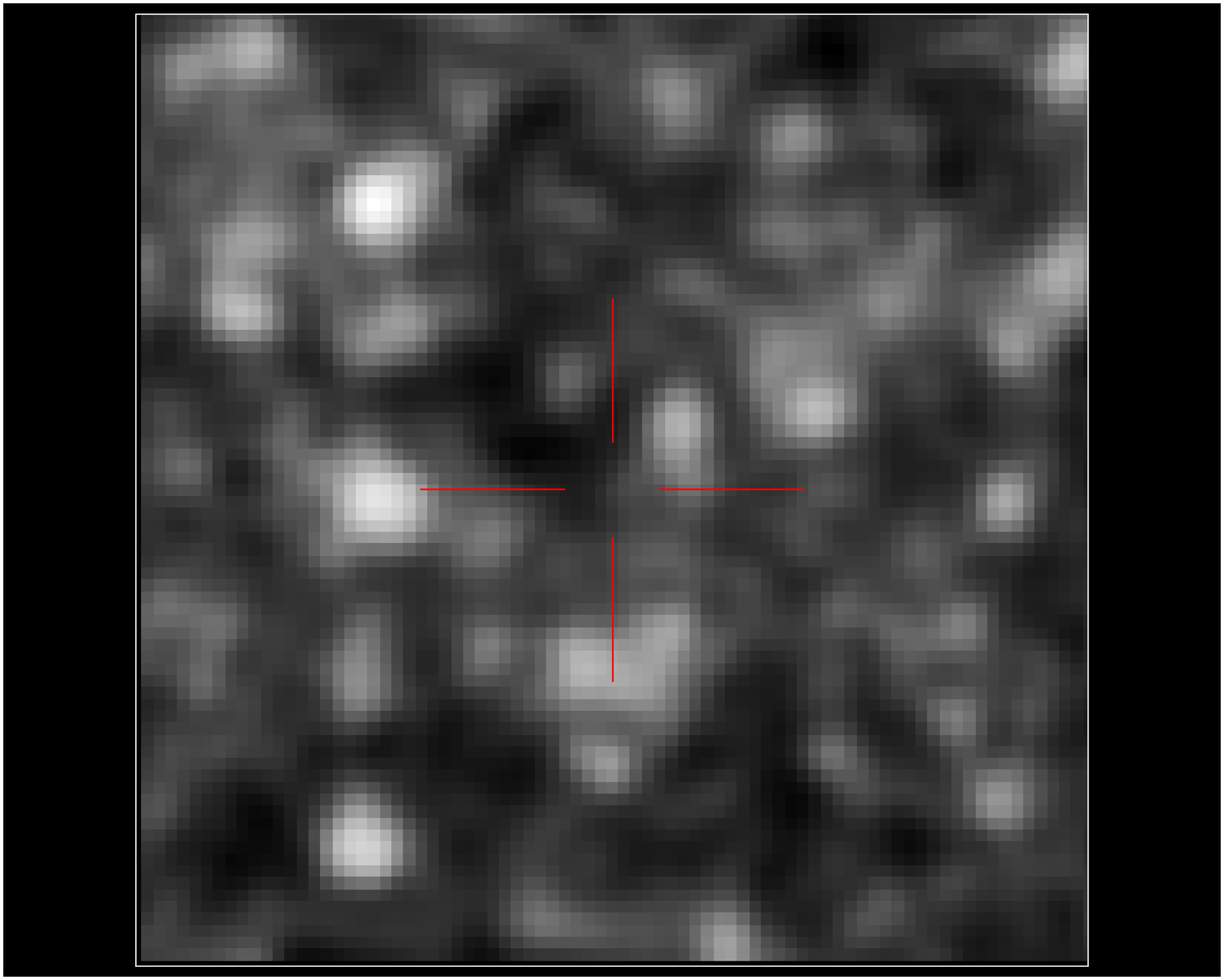}}
  \subfloat[G230.97+32.31]{\includegraphics[width=0.25\textwidth]{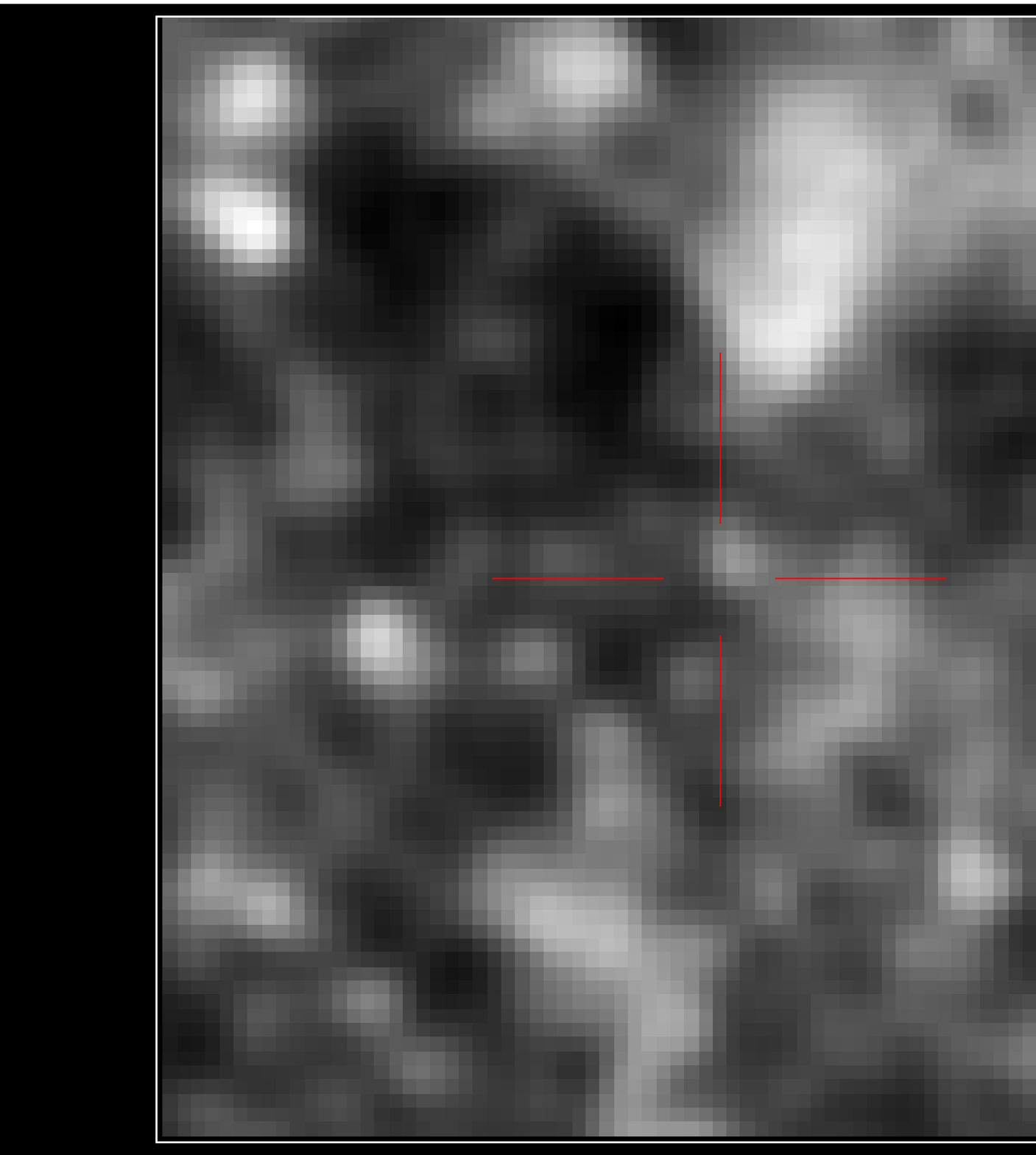}}
  \\
  
  \subfloat[G231.25+32.05]{\includegraphics[width=0.25\textwidth]{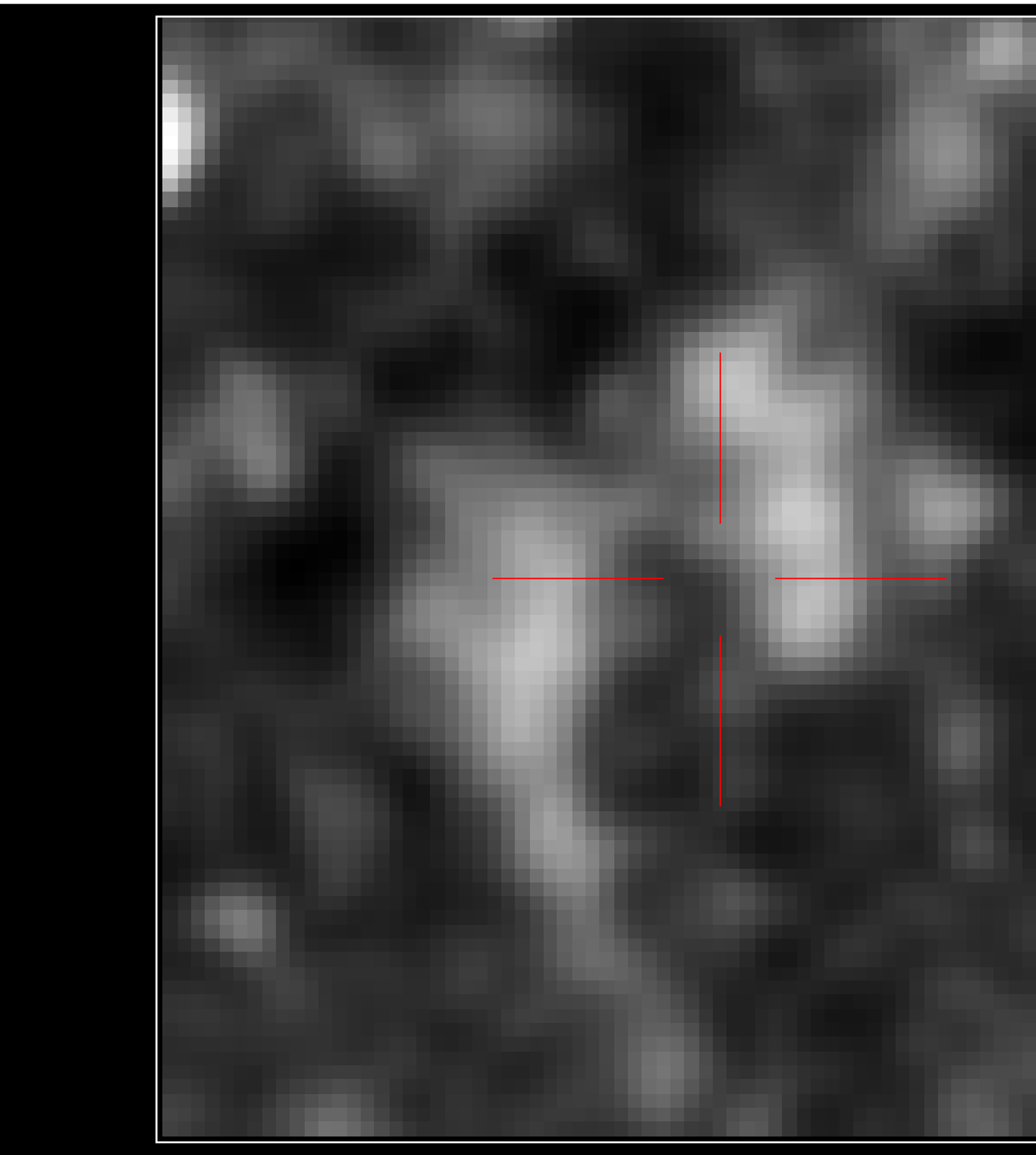}}
  \subfloat[G231.38+32.24]{\includegraphics[width=0.25\textwidth]{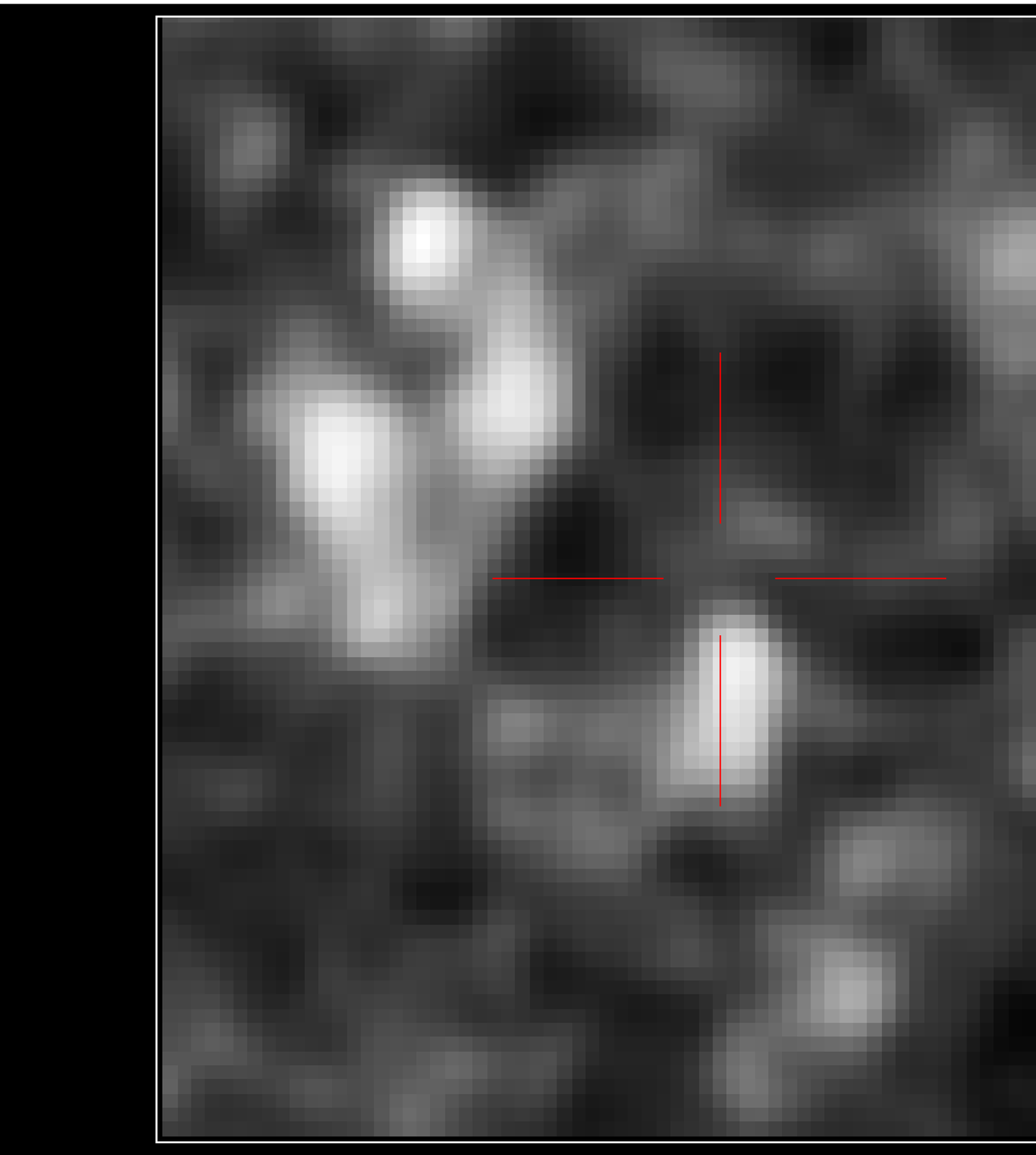}}
  \subfloat[G231.43+32.10]{\includegraphics[width=0.25\textwidth]{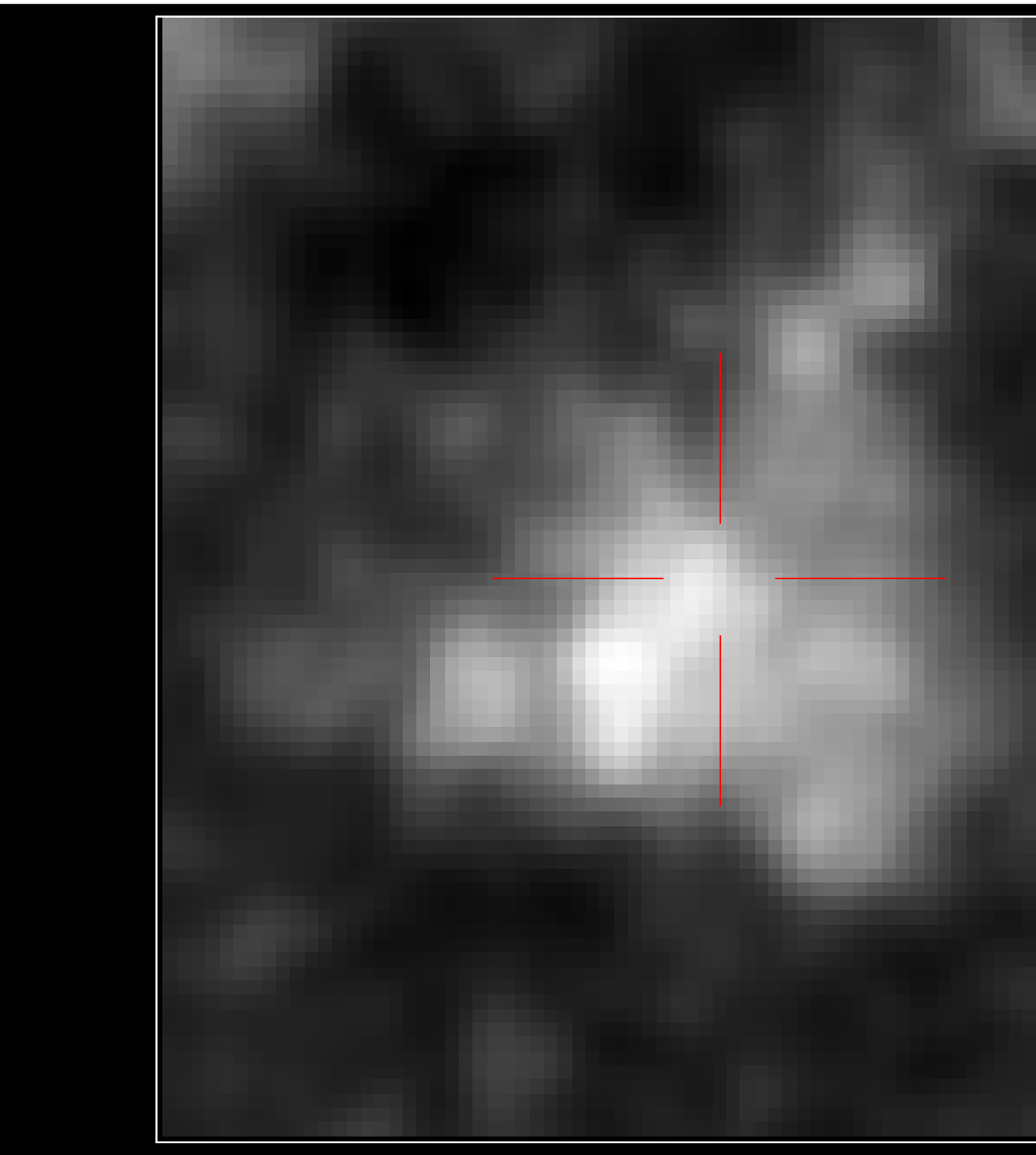}}
  \subfloat{\includegraphics[width=0.25\textwidth]{Figs/Poststamps/blank.eps}}
    \caption{SPIRE images at 350 $\mu$m showing postage stamps of the \emph{Planck} ERCSC sources at 857 GHz in the H-ATLAS GAMA-09 field. The images are 400 arcsec wide and the red crosses indicate the ERCSC position for each source. The full ERCSC name of the objects has been abbreviated in the captions for conciseness.}
  \label{fig:poststamps_GAMA-09}
\end{figure*}

\begin{figure*}
  \centering
  \subfloat[G263.84+57.55]{\includegraphics[width=0.25\textwidth]{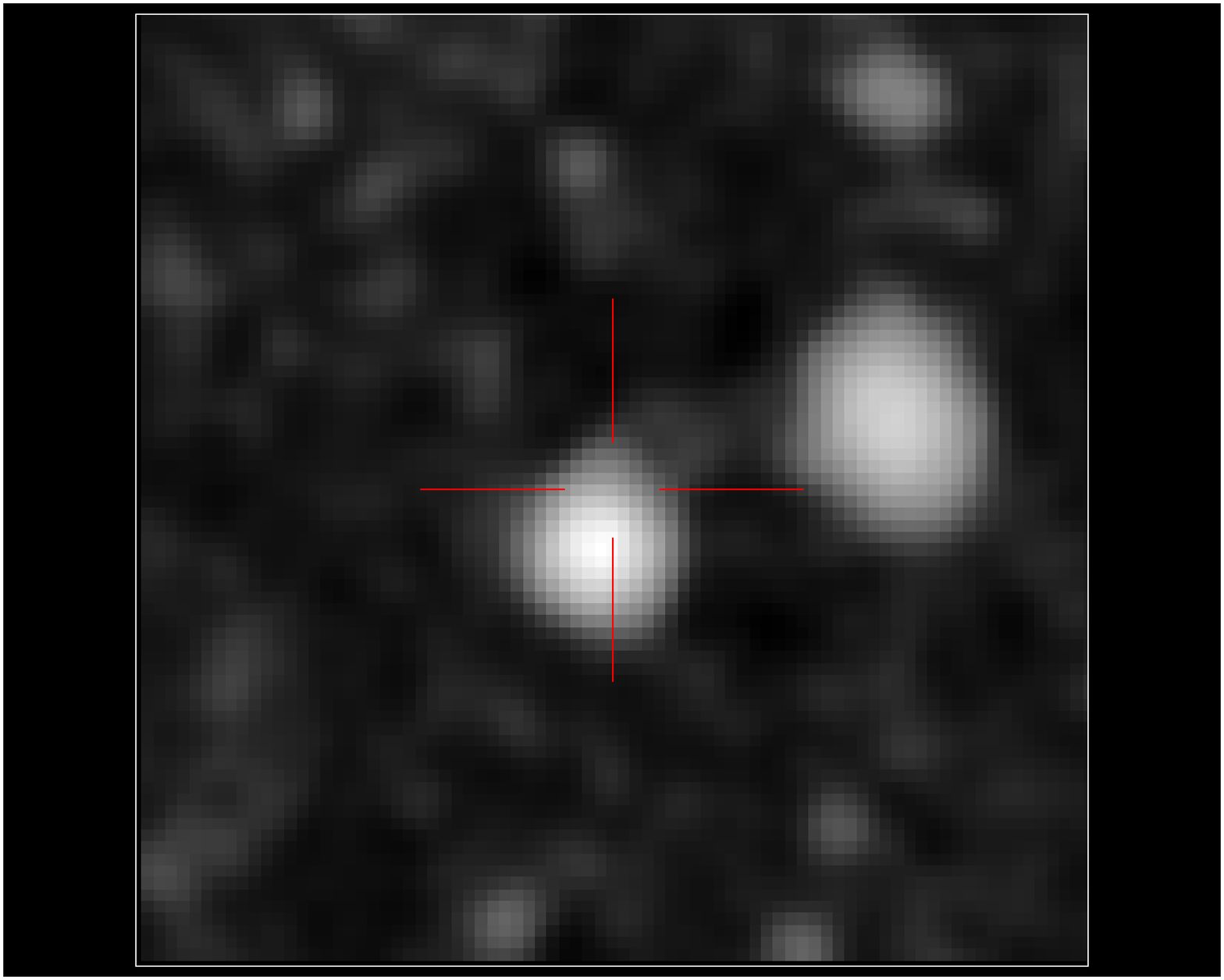}}
  \subfloat[G266.26+58.99]{\includegraphics[width=0.25\textwidth]{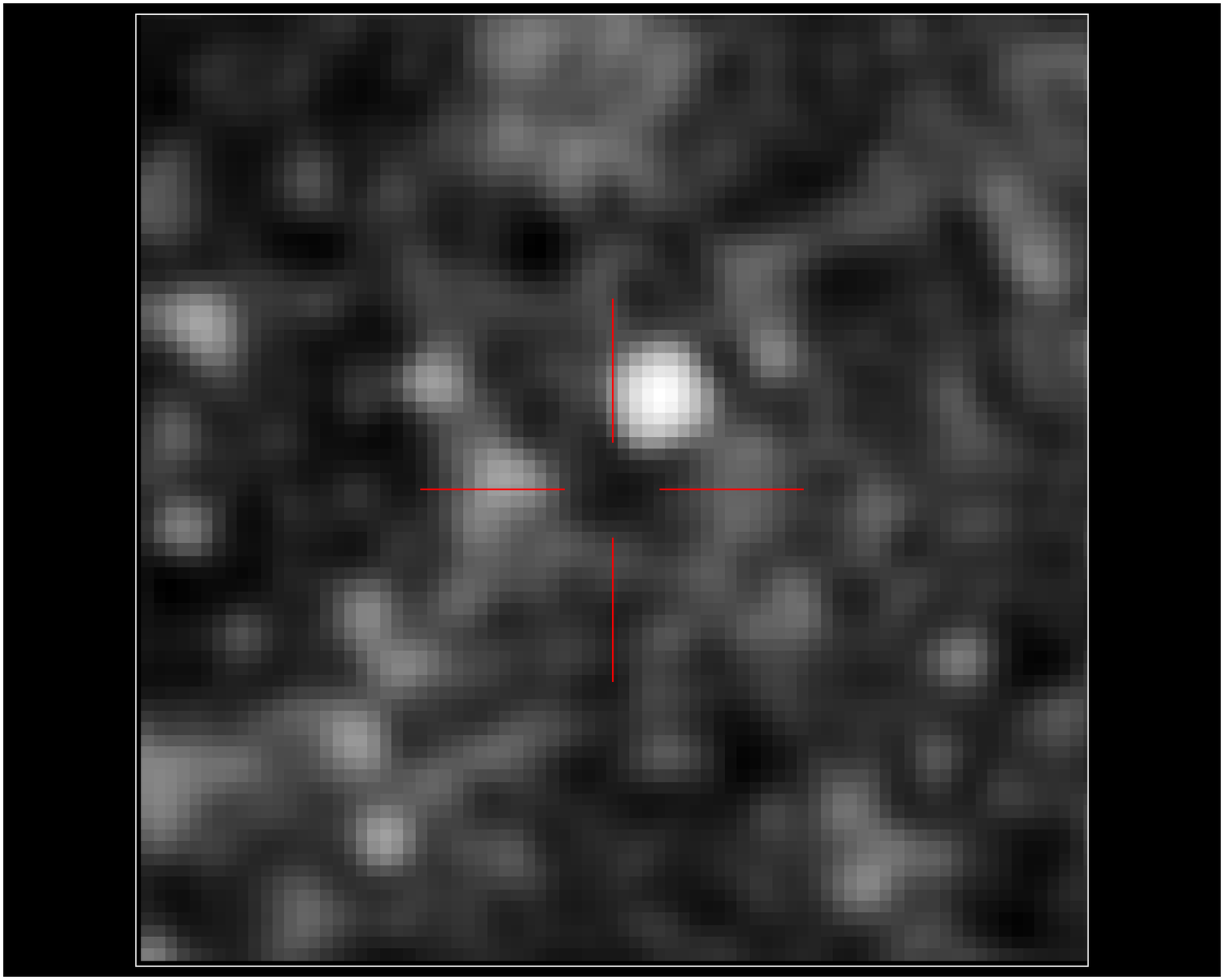}}
  \subfloat[G270.59+58.52]{\includegraphics[width=0.25\textwidth]{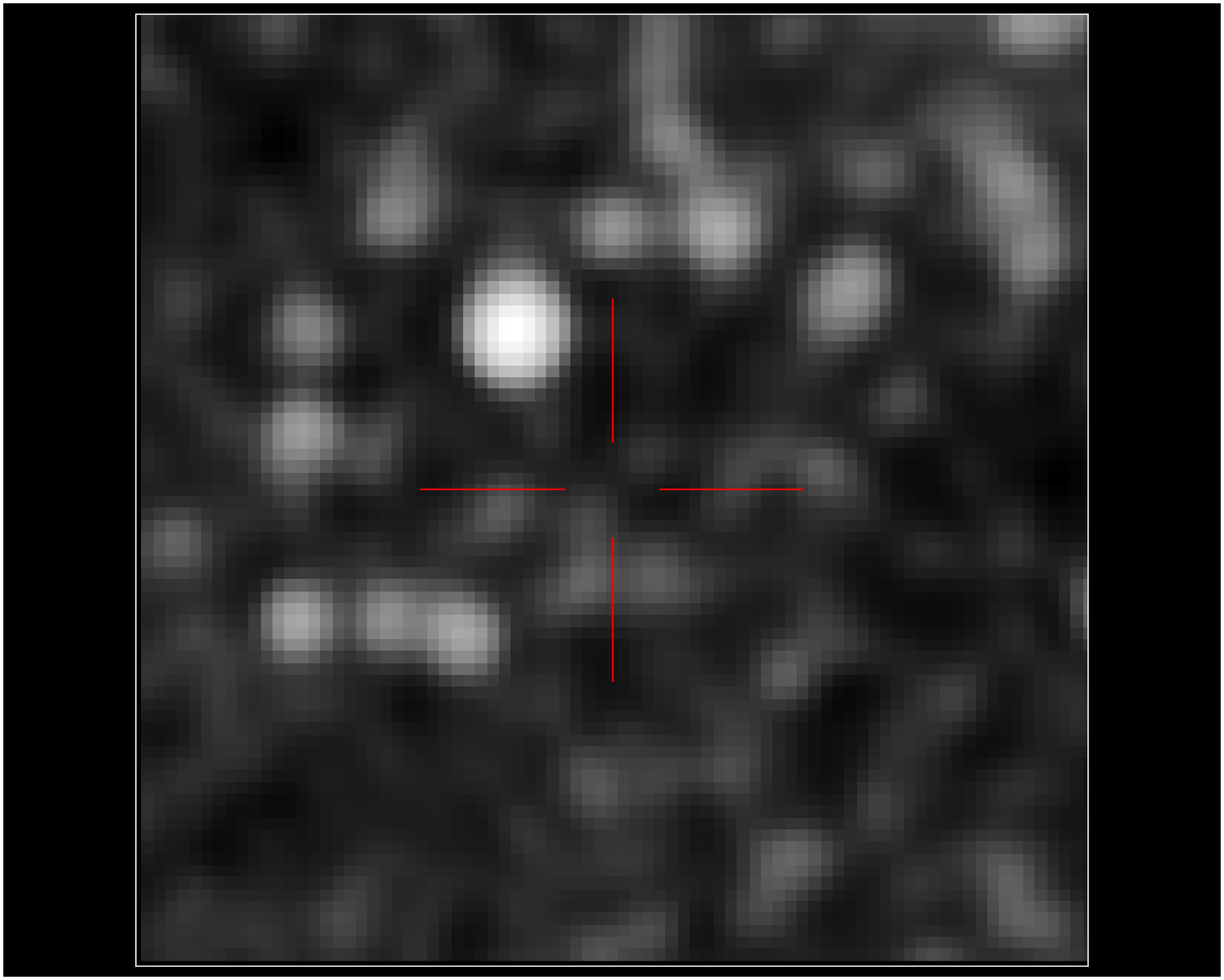}}
  \subfloat[G274.04+60.90]{\includegraphics[width=0.25\textwidth]{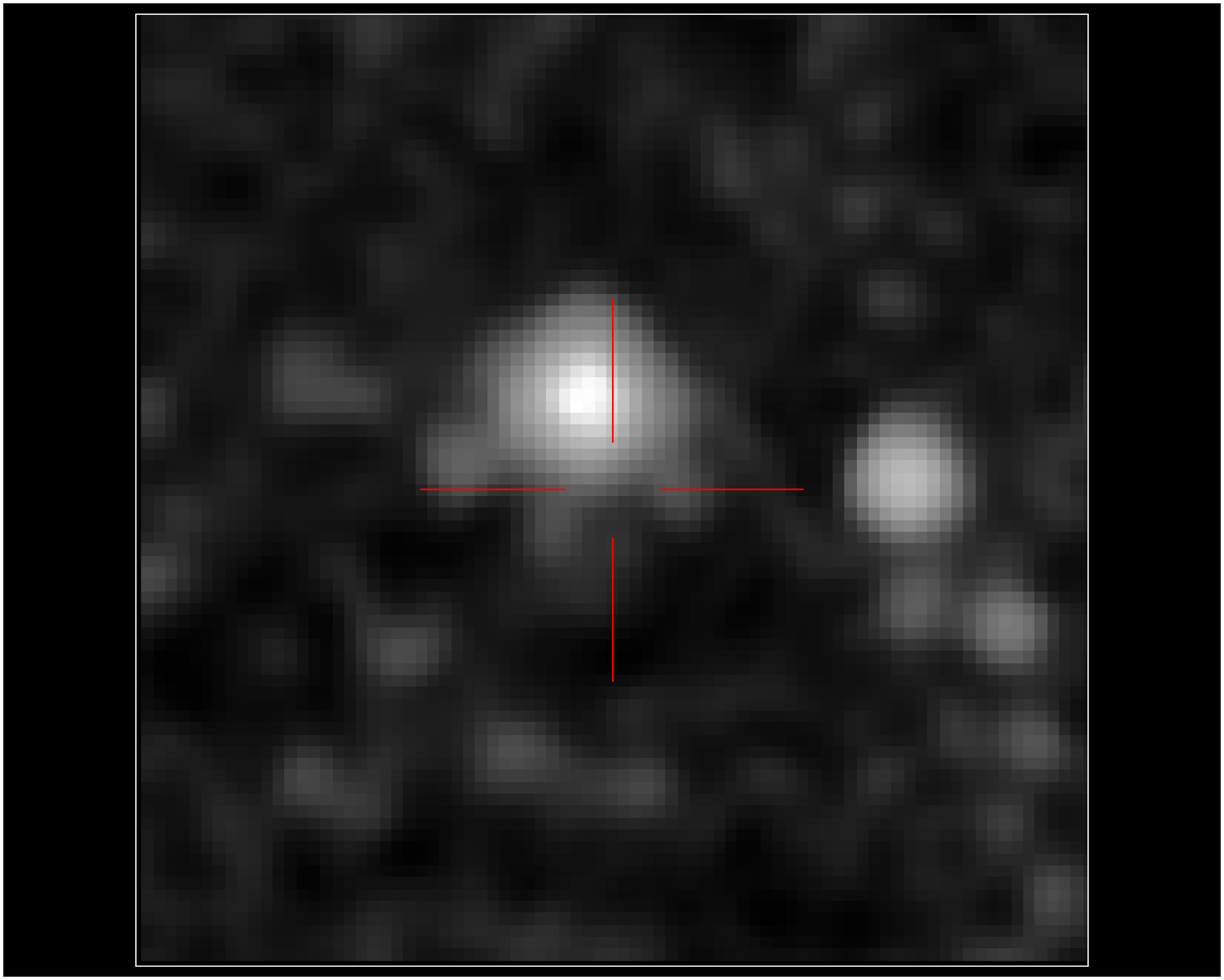}}
   \\
  \subfloat[G277.37+59.21]{\includegraphics[width=0.25\textwidth]{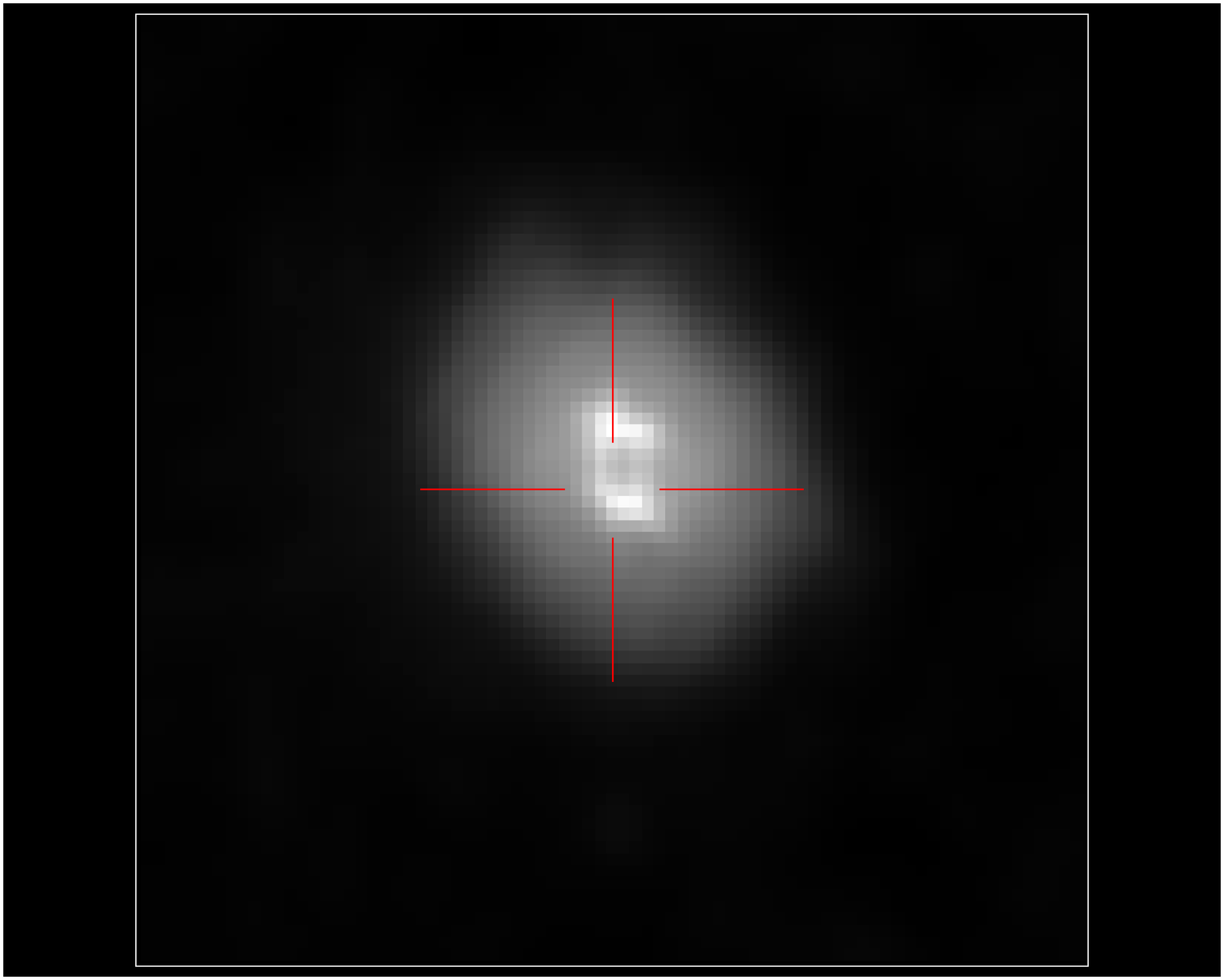}}
  \subfloat{\includegraphics[width=0.25\textwidth]{Figs/Poststamps/blank.eps}}
  \subfloat{\includegraphics[width=0.25\textwidth]{Figs/Poststamps/blank.eps}}
  \subfloat{\includegraphics[width=0.25\textwidth]{Figs/Poststamps/blank.eps}}
  \caption{SPIRE images at 350 $\mu$m showing postage stamps of the \emph{Planck} ERCSC sources at 857 GHz in the H-ATLAS GAMA-12 field. The images are 400 arcsec wide and the red crosses indicate the ERCSC position for each source.}
  \label{fig:poststamps_GAMA-12}
\end{figure*}

\begin{figure*}
  \centering
  
  \subfloat[G345.11+54.84]{\includegraphics[width=0.25\textwidth]{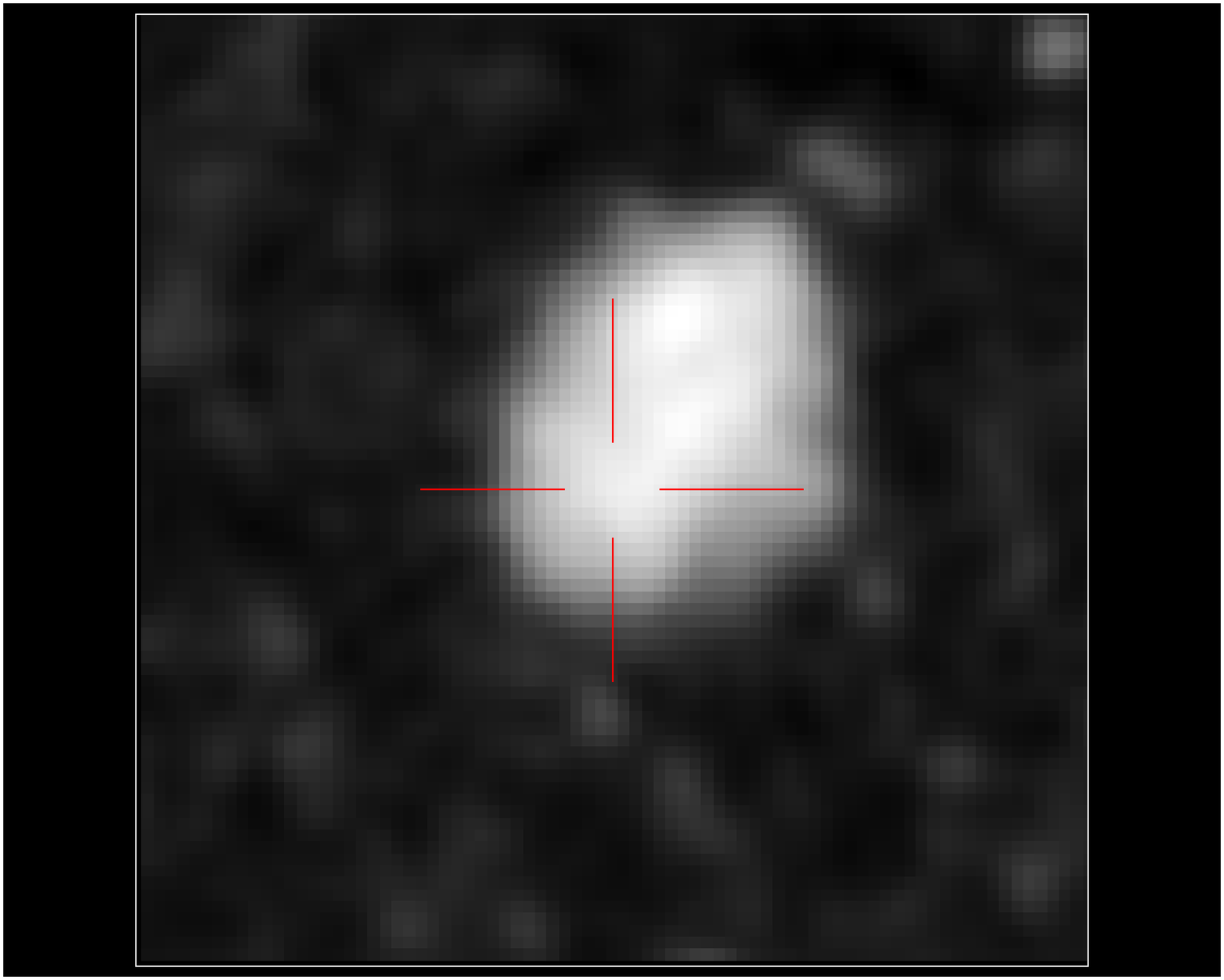}}
  \subfloat[G347.77+56.35]{\includegraphics[width=0.25\textwidth]{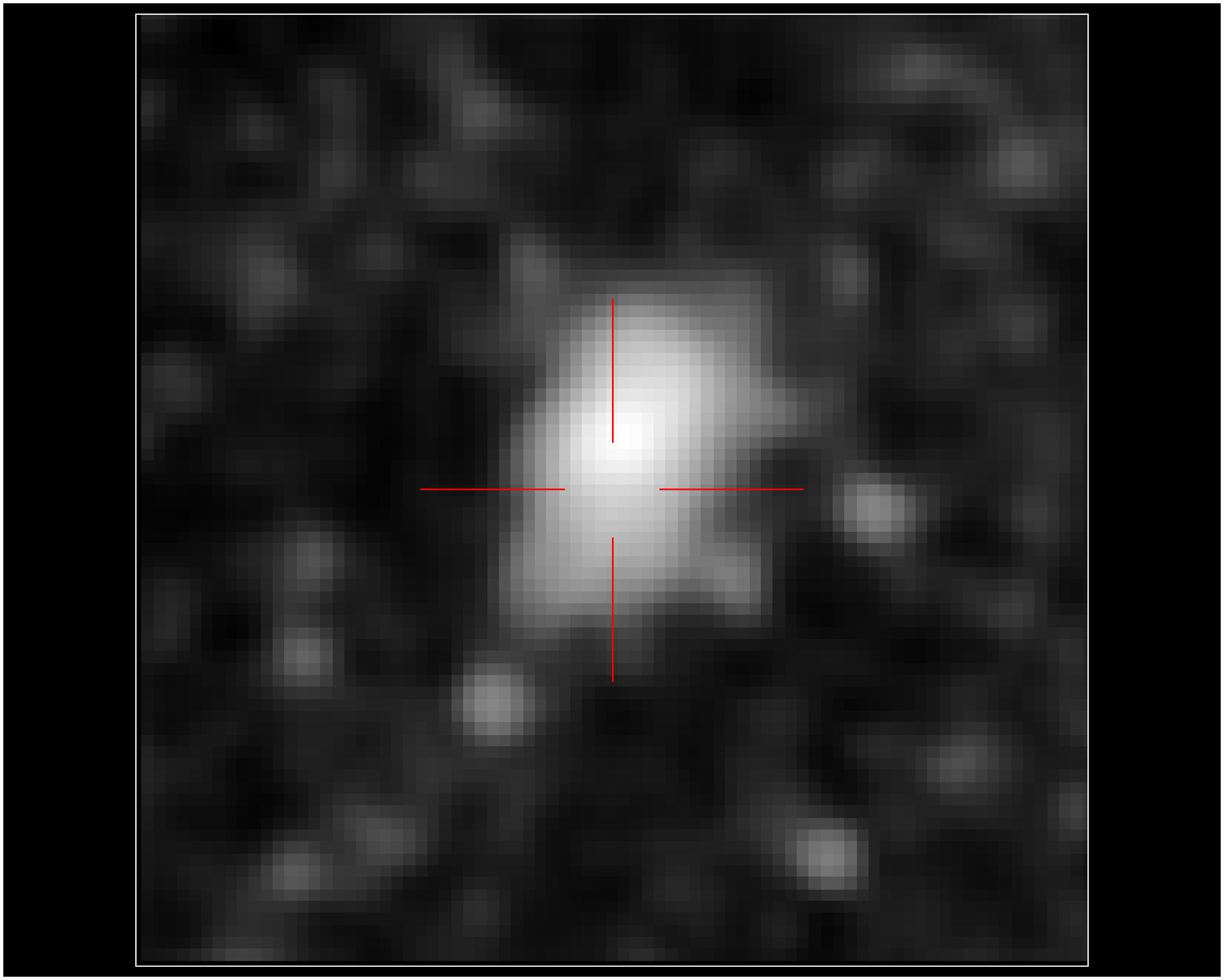}}
  \subfloat[G350.46+51.85]{\includegraphics[width=0.25\textwidth]{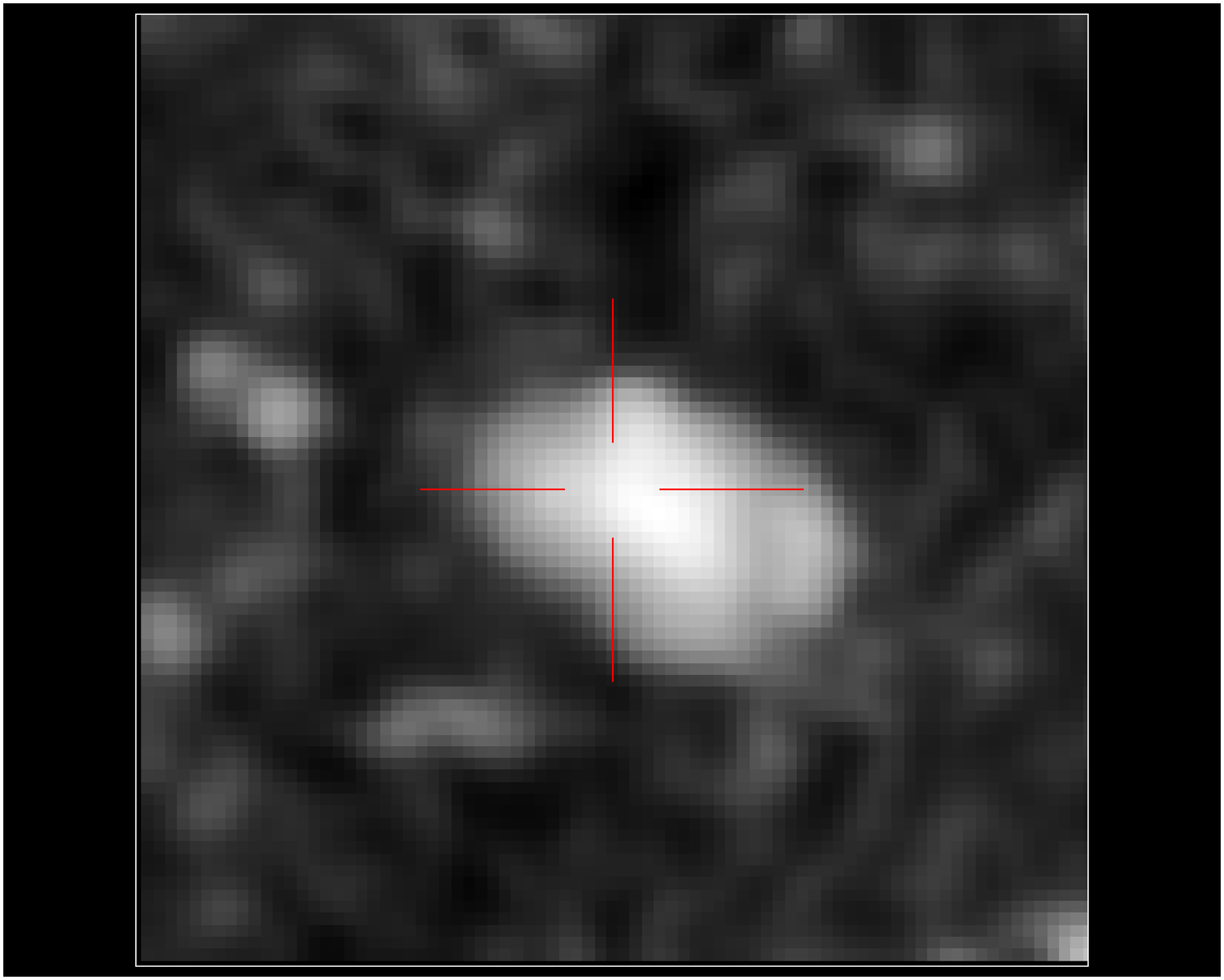}}
  \subfloat[G351.01+52.11]{\includegraphics[width=0.25\textwidth]{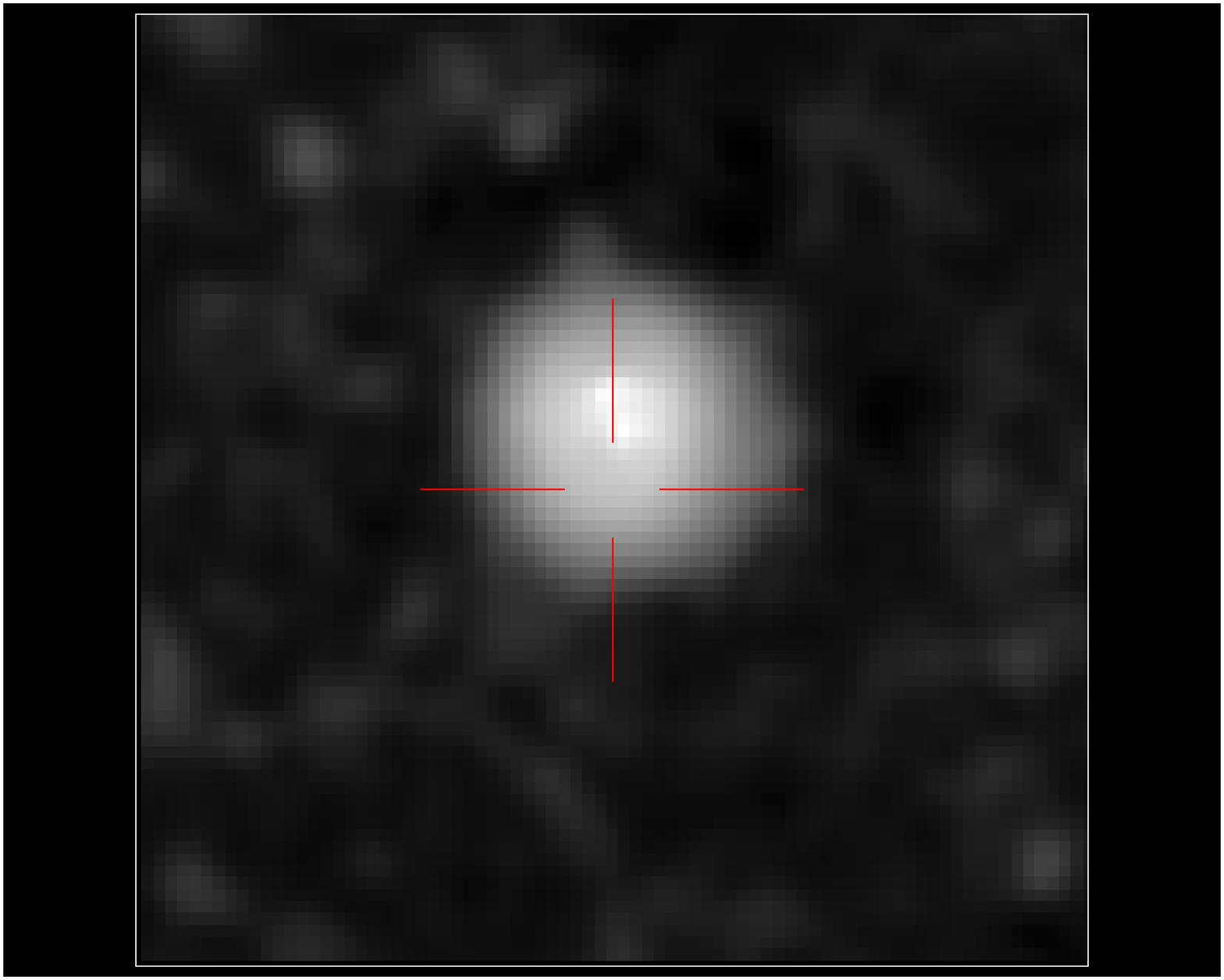}}
  \\
  
  \subfloat[G351.22+51.97]{\includegraphics[width=0.25\textwidth]{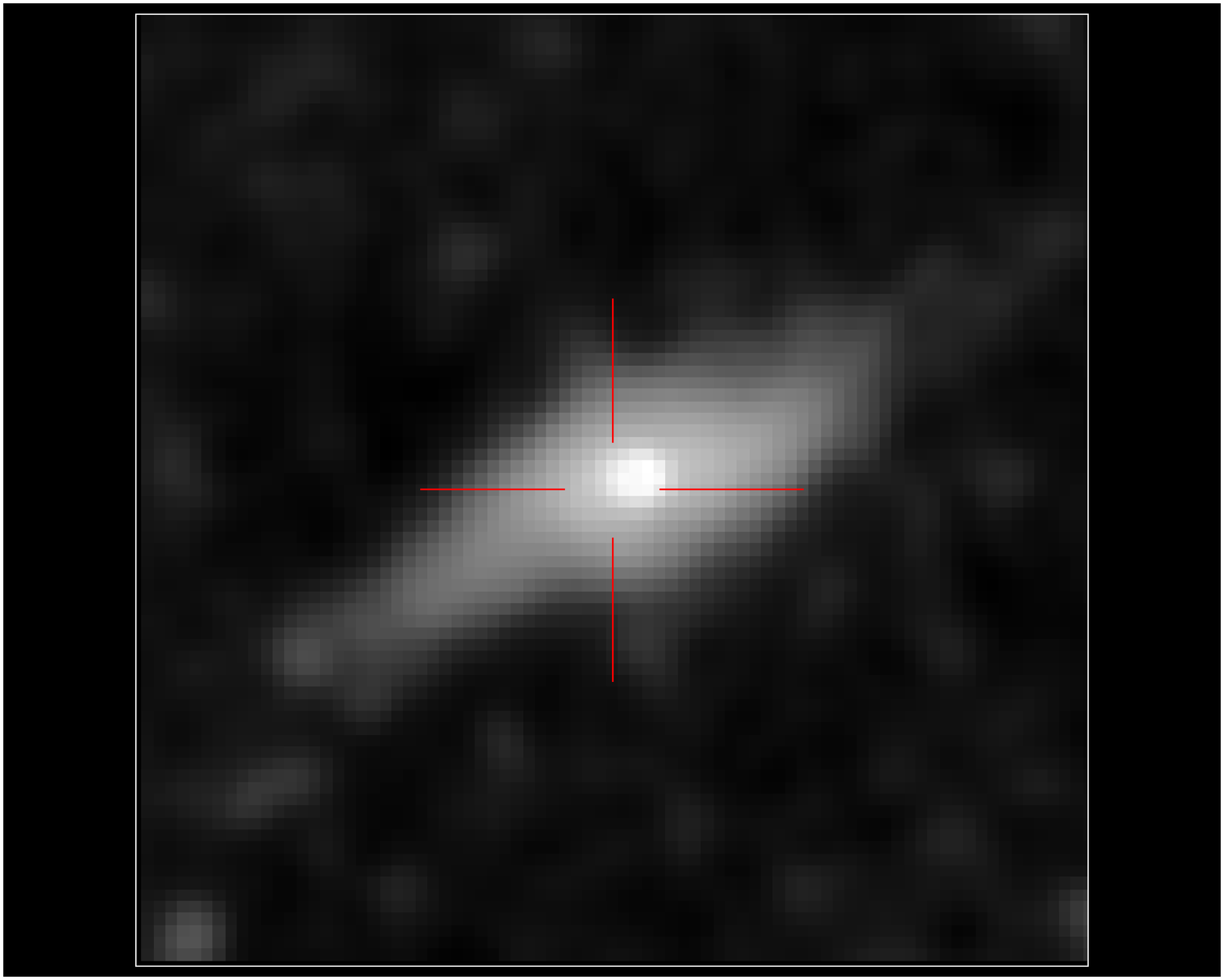}}
  \subfloat[G353.15+54.45]{\includegraphics[width=0.25\textwidth]{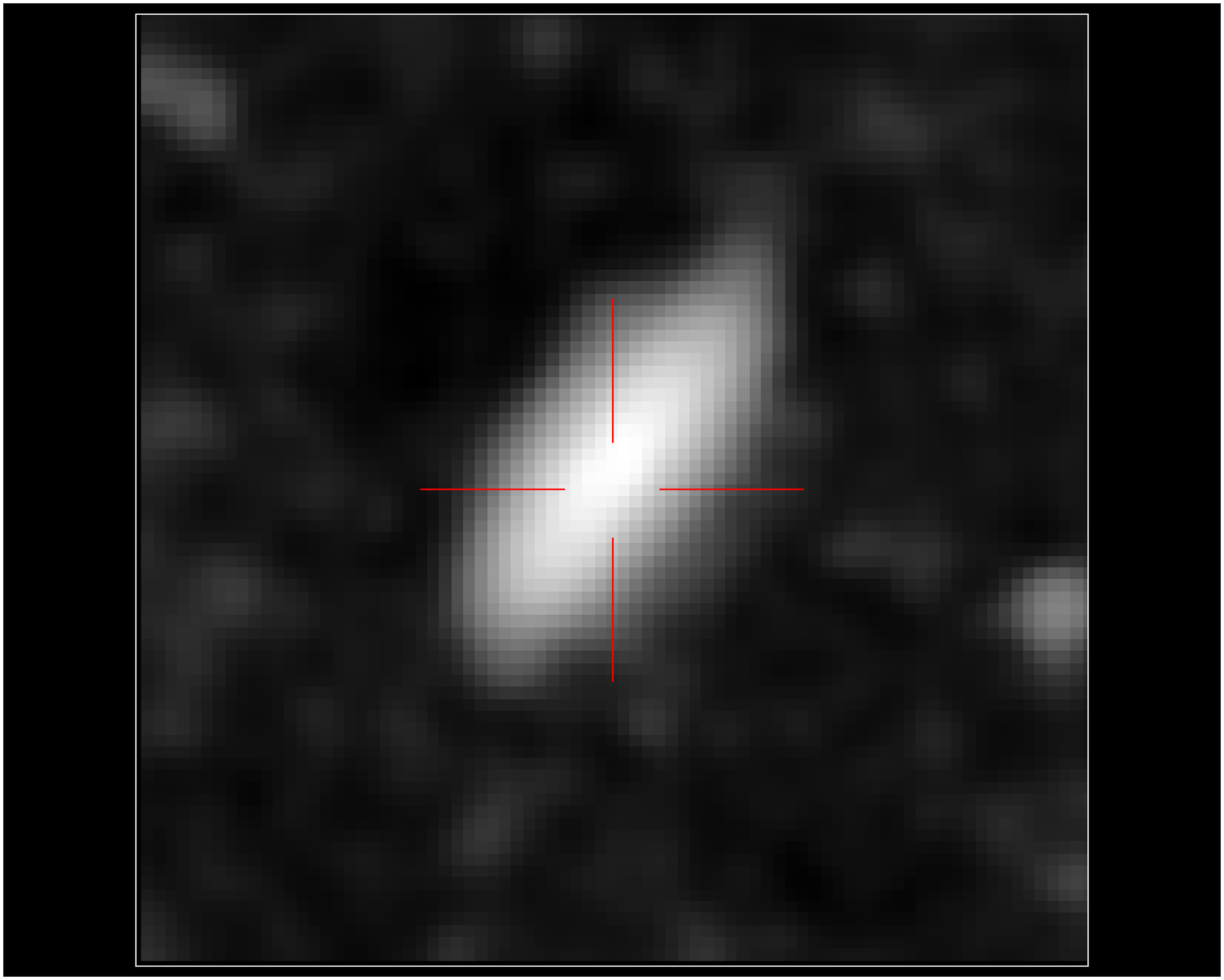}}
  \subfloat[G354.50+52.84]{\includegraphics[width=0.25\textwidth]{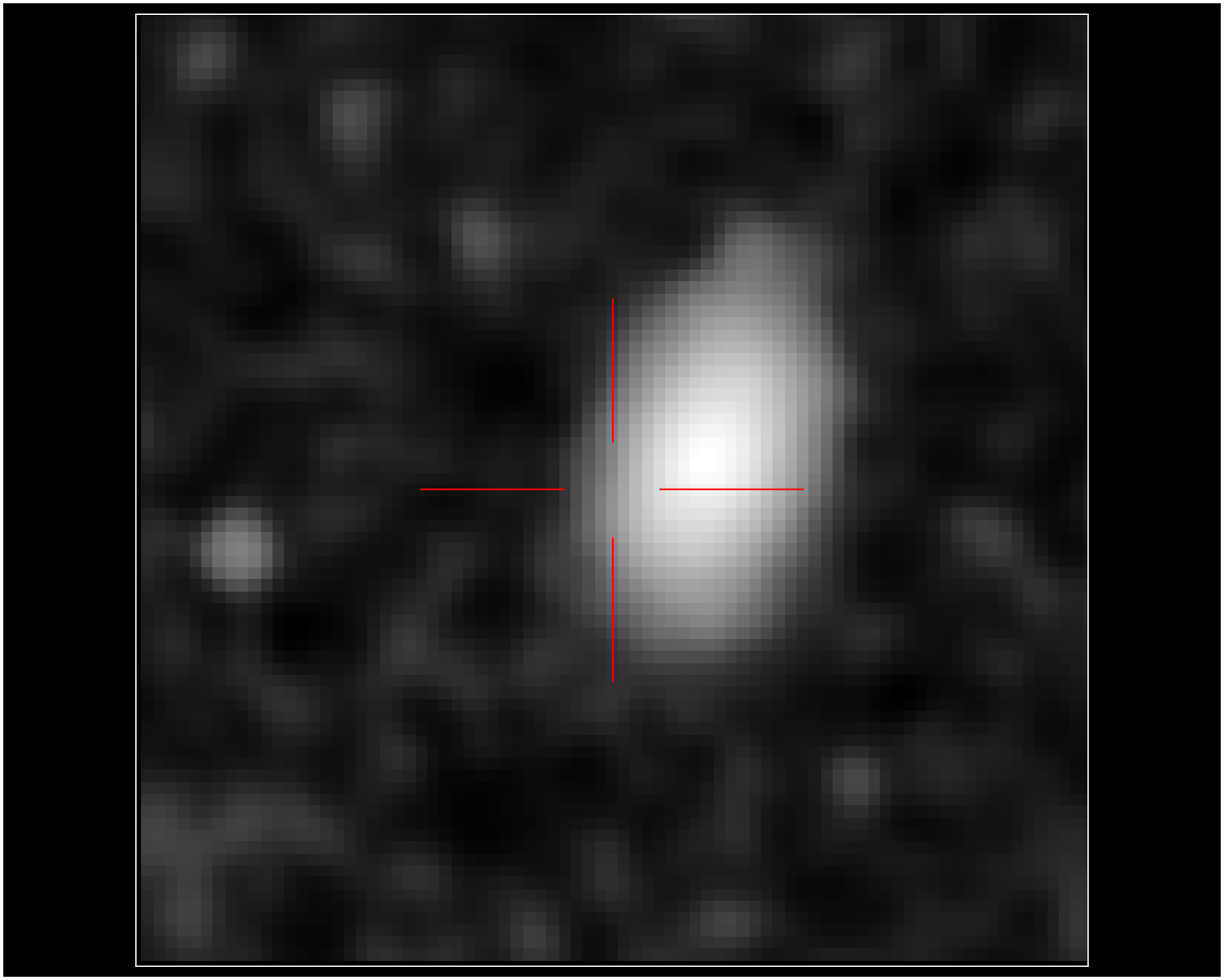}}
  \subfloat[G354.50+52.84]{\includegraphics[width=0.25\textwidth]{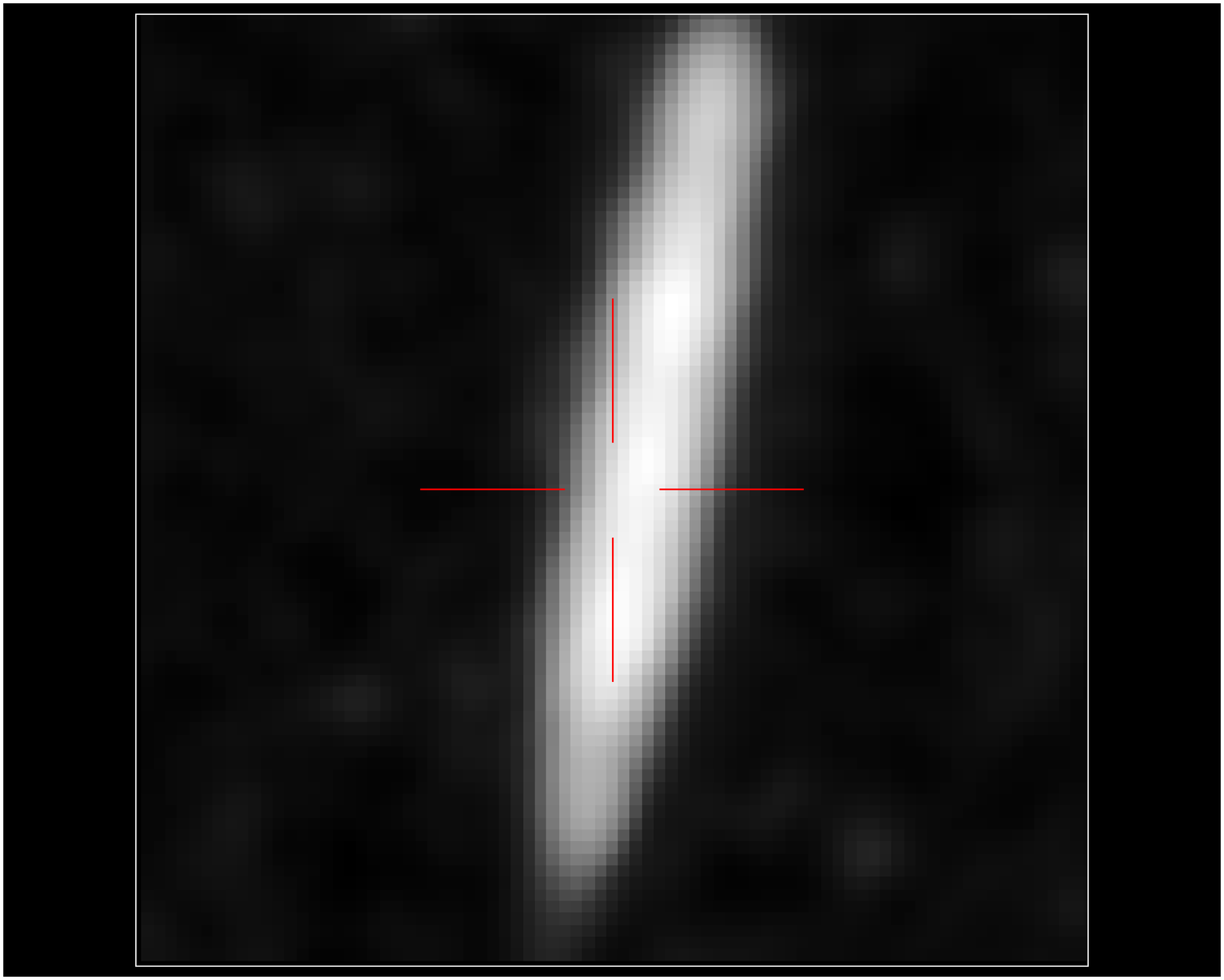}}
    \caption{SPIRE images at 350 $\mu$m showing postage stamps of the \emph{Planck} ERCSC sources at 857 GHz in the H-ATLAS GAMA-15 field. The images are 400 arcsec wide and the red crosses indicate the ERCSC position for each source.}
  \label{fig:poststamps_GAMA-15}
\end{figure*}

The two highest frequency channels of the \emph{Planck} HFI practically overlap with the two lower frequency bands of \emph{Herschel} SPIRE. The 350 $\mu$m band almost coincides in central wavelength and bandwidth with the \emph{Planck} HFI 857 GHz channel. The 500 $\mu$m band and the \emph{Planck} HFI 545 GHz channel do not coincide exactly, but are close enough to consider a cross-check. The H-ATLAS Phase 1 covers $134.55\,\hbox{deg}^2$ divided into three regions (GAMA-09, GAMA-12 and GAMA-15).

\subsection{H-ATLAS $350\,\mu$m counterparts of ERCSC 857\,GHz sources} \label{sec:350-857}

Within the H-ATLAS Phase 1 fields there are 28 ERCSC sources detected by \emph{Planck} at 857\,GHz. Among these, there are no clusters of galaxies (detected through the Sunyaev-Zel'dovich effect) or cold cores. Figures~\ref{fig:poststamps_GAMA-09}, ~\ref{fig:poststamps_GAMA-12} and ~\ref{fig:poststamps_GAMA-15} show 
postage stamp images of these 28 sources as viewed by \emph{Herschel} at 350 $\mu$m in the GAMA-09, GAMA-12 and GAMA-15 fields, respectively. By inspecting the SPIRE images around the positions of the sources we find several different situations.

As many as 16 out of the 28 ERCSC objects do not have a consistent H-ATLAS counterpart, because H-ATLAS sources within the Planck beam are too faint to explain the flux density measured by \emph{Planck}. Almost all (15) of them are flagged as extended and/or have a relatively high ($\ge 0.125$) cirrus flag in the ERCSC (11 have both properties). All but \object{G266.26+58.99} (object (b) in Figure~\ref{fig:poststamps_GAMA-12} which, incidentally, is not flagged as extended and has a relatively low cirrus flag) are in the GAMA-09 field, which is more contaminated by Galactic emission than the other two GAMA fields. Fig.~\ref{fig:iris} shows the positions of the ERCSC sources around the GAMA-09 field superimposed to the IRIS 100 micron map \citep{IRIS}. The high correlation between ERCSC sources and the IRIS map reinforces the idea that the GAMA-09 ERCSC sources are likely to be mostly related to Galactic cirrus.

As for the other 12 sources, one (\object{G263.84+57.55}, the object (a) in Figure~\ref{fig:poststamps_GAMA-12}) is clearly resolved by \emph{Herschel} into two relatively bright sources, the galaxy pair \object{KPG289} \citep{kar76} formed by the galaxies \object{NGC3719} and \object{NGC3720}, each with flux density $>250$ mJy at 350 $\mu$m, while another object (\object{G270.59+58.52}, shown in panel (c) of Figure~\ref{fig:poststamps_GAMA-12}) is resolved into an unusual condensation of low flux, probably high-redshift point sources. We will discuss this source in more detail in \S\,\ref{sec:proto}. Finally, 10 ERCSC sources can be clearly identified with single bright H-ATLAS sources at low redshift, including the edge-on spiral \object{NGC5746} (object (h) in Figure~\ref{fig:poststamps_GAMA-15}) the spirals \object{NGC5690}, \object{NGC5705} and \object{NGC4030}, the peculiar asymmetric galaxy  \object{NGC5713} \citep{dale12}, in addition to the the above mentioned pair \object{NGC3719} and \object{NGC3720}.

Table~\ref{table2} lists the 28 sources with their 857 GHz flux densities taken from the ERCSC \citep{ERCSC_cat}. The table gives the ERCSC source name, the RA and Dec coordinates, the flux densities and associated errors, and the \textsc{EXTENDED} and \textsc{CIRRUS} flags. Column 9 gives the $350\,\mu$m flux densities of the brightest H-ATLAS sources found inside the \emph{Planck} beam\footnote{Unless otherwise stated, the beam is a circle whose radius is $r=\mathrm{FWHM}/2\sqrt{2\log 2}$.}, provided that they have $S_{350}\ge 250\,$mJy. For these sources we give, in column 11, the spectroscopic or photometric redshifts taken from the H-ATLAS catalog (Dunne et al., in preparation). The positions of the 28 sources in the ATLAS GAMA09, GAMA-12 and GAMA-15 fields are shown in Figures~\ref{fig:GAMA-09}, \ref{fig:GAMA-12} and \ref{fig:GAMA-15} respectively.

\subsection{H-ATLAS $500\,\mu$m counterparts of ERCSC 545\,GHz sources}

Turning now to longer wavelengths,
all but one (\object{PLCKERC545 G230.17+32.05}) of the 14 ERCSC sources detected by \emph{Planck} at 545\,GHz that lie in the H-ATLAS Phase 1 fields are among the 28 objects described in \S\,\ref{sec:350-857}. The object not detected by \emph{Planck} at 857 GHz is in the GAMA09 region and has a high CIRRUS flag. Like at 857 GHz, more than half of the sources (8 out of 14, 7 of them being in the GAMA09 field) do not have a plausible H-ATLAS counterpart. Table~\ref{table3} lists the 14 sources, giving the H-ATLAS ID and redshifts for the six sources which have a 500 $\mu$m counterpart with $S_{500}\gsim 250\,$mJy.

\section{Photometry} \label{sec:photo}

The comparison of the flux density estimations of the ERCSC and the SPIRE Phase 1 catalogues is not straightforward. In order to correctly compare ERCSC and SPIRE photometric estimations it is necessary to take into account that:
\begin{itemize}
\item \emph{Herschel} has better angular resolution than \emph{Planck}. It is possible that an ERCSC source can be resolved into several sources by \emph{Herschel}.
\item The wavelengths of the \emph{Planck} bands do not correspond exactly to the wavelenghts of their \emph{Herschel} counterparts.
\item Both catalogues have been obtained by using different detection and photometry extraction algorithms.
\end{itemize}

The effect of the different angular resolutions can be corrected, at least to first order, by integrating the SPIRE flux densities over the larger \emph{Planck} beam area and weighting by the \emph{Planck} beam response. The effect of the different central wavelenghts can be taken into account by means of SED colour correction, as will be described in \S\ \ref{sec:500vs545}. 
A review of the technical details of the ERCSC and SPIRE flux density estimates used in the catalogues is beyond the scope of this work. ERCSC photometry is described in \cite{ERCSC,expla}; SPIRE photometry is described in \cite{rigby11}; both catalogues have passed strict internal and external validation. For pointlike sources, we assume that the flux density estimates that are given in both catalogues are accurate, within the calibration uncertainties of their experiments.
However, it must be noted that the different ways in which the photometry of extended sources is done in the ERCSC and in the ATLAS Phase 1 catalogues can affect the comparison.
The ATLAS catalogues use  aperture photometry scaled to the optical size of the sources \citep{rigby11}. The \emph{Planck} ERCSC contains several types of photometric measurements for each source; the default aperture photometry information is listed in the ERCSC `FLUX' column, but other photometric measurements may be more appropriate in certain circumstances. As described in the \emph{Planck} Explanatory Supplement to the ERCSC \citep{expla}, for extended sources it may be better to use the Gaussian-fit photometry listed in the ERCSC `GAUFLUX' column instead of a fixed aperture photometry. Since a significant fraction of the sources in our sample at 857 GHz are flagged as extended in the ERCSC, we have compared the H-ATLAS flux densities to both the fixed aperture photometry (`FLUX') and the Gaussian-fit (`GAUFLUX') ERCSC flux densities. We find that at the lowest fluxes FLUX and GAUFLUX work similarly well, while at bright fluxes GAUFLUX is more consistent with \emph{Herschel} photometry (may be because many bright ERCSC sources are extended). Therefore, in the rest of this paper, we will use GAUFLUX when referring to \emph{Planck} photometry.

\subsection{ERCSC-857 GHz and H-ATLAS $350 \mu m$ flux densities}

According to the technical specifications of their respective instruments, the \emph{Planck}/HFI 857 GHz  and the \emph{Herschel}/SPIRE 350 $\mu$m channels have almost exactly the same central wavelength and roughly the same bandwidth. This makes it easy to directly compare the flux density estimations of the ERCSC sources present in the H-ATLAS fields. In order to take into account the different angular resolutions of \emph{Herschel} and \emph{Planck}, we have calculated an effective \emph{Herschel} 350 $\mu$m flux density by summing up the flux densities, corrected for the effect of the \emph{Planck} beam, of the sources listed in the H-ATLAS Phase 1 catalogue around the ERCSC positions. We have assumed a circular, Gaussian beam of $\hbox{FWHM} = 4.23$ arcmin \citep{ERCSC}.

\begin{figure}
  \resizebox{\hsize}{!}{\includegraphics{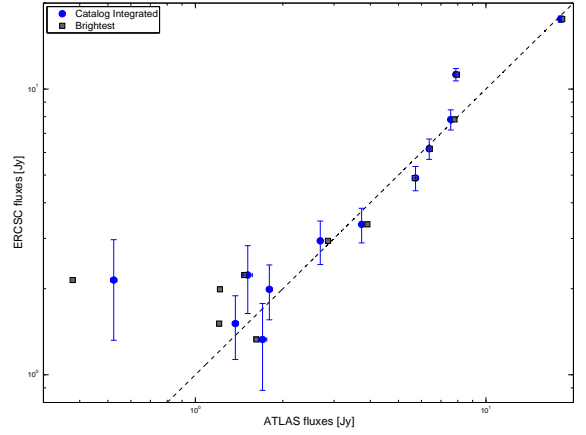}}
  \caption{ERCSC 857 GHz flux densities compared with the 350 $\mu$m flux densities of the brightest H-ATLAS sources inside the \emph{Planck} beam (squares). The dashed line indicates the $S_{\mathrm{ATLAS}}=S_{\mathrm{ERCSC}}$ identity. The filled circles show the summed flux densities, weighted with a Gaussian beam centered on the ERCSC position and with $\hbox{FWHM}=4.23$ arcmin \citep{ERCSC}. Only the ERCSC sources with at least one \emph{Herschel} counterpart with flux density $S_{\rm {\tiny ATLAS}}>0.25$ Jy are shown on this plot (see text for further details). The outlier with ERCSC flux density $\sim 10$ Jy is the edge-on spiral galaxy NGC5746 (see discussion in the main text). The outlier with ERCSC flux density $\sim 2$ Jy is the G12H29 source to be discussed in \S\ 5.}
  \label{fig_flux_comparison_350vs857}
\end{figure}

Fig.~\ref{fig_flux_comparison_350vs857} shows the \emph{Planck} 857 GHz flux densities compared to \emph{Herschel} $350 \mu m$ flux densities for those ERCSC sources that have at least one $S_{\rm {\tiny ATLAS}} \geq 250$ mJy. All these sources lie in the GAMA-12 or GAMA-15 fields.  
According to \cite{expla}, the ERCSC flux densities below $\simeq 1.3\,$Jy are boosted by the well known selection bias (sources sitting on top of positive noise plus confusion plus Galactic emission fluctuations that dominate the contribution to the measured flux densities are more likely to be detected). 
Leaving aside the large edge-on spiral NGC5746, which will be discussed below, for sources with H-ATLAS $350\mu$m flux densities $\ge 1.5\,$Jy we see a good agreement between \emph{Herschel} and \emph{Planck} flux densities. We find $\langle S_{\rm {\tiny ATLAS}}-S_{\rm {\tiny ERCSC}}\rangle\simeq 0.1\,$Jy. The relative flux density difference for the same sources, defined as
\begin{equation}
\epsilon = 100 \times \langle \frac{S_{\rm {\tiny ATLAS}}-S_{\rm {\tiny ERCSC}}}{S_{\rm {\tiny ATLAS}}} \rangle,
\end{equation}
\noindent
is $\epsilon = -3\%$, smaller than the calibration uncertainty of both \emph{Herschel} and \emph{Planck}. However, the relative difference of the individual sources shows a large ($\sim 20\%$) scatter, probably due to the small size of the sample.
The mean difference between ERCSC and H-ATLAS positions for sources brighter than 1.5 Jy is 0.42 arcmin, with a dispersion of 0.17 arcmin. For the assumed beam shape, this may account for a $\simeq 5\%$ underestimate of ERCSC flux densities.

The two remarkable outliers in Figure~\ref{fig_flux_comparison_350vs857} are the sources $\#18$ and $\#28$ in Table~\ref{table2}.
The source $\#18$, \object{G270.59+58.52} (object (c) in Figure~\ref{fig:poststamps_GAMA-12}), flagged as extended in the ERCSC and with $S_{\rm {\tiny ATLAS}}=352\,$ mJy will be discussed in \S\,\ref{sec:proto}.
The source  $\#28$ (object (h) in Figure~\ref{fig:poststamps_GAMA-15}) is identified as NGC5746, a large edge-on spiral that is clearly resolved as a very extended  source by SPIRE but is not flagged as extended by the ERCSC. We believe that the discrepancy between the flux densities for this object reported by H-ATLAS and the ERCSC is due to the very different angular resolution of \emph{Herschel} and \emph{Planck} and by the different way in which background subtraction has been performed by the catalog making pipelines of the two experiments. In particular, if aperture photometry is applied to the raw (non background subtracted) SPIRE 350 micron map, a flux density of $\sim 10$ Jy is obtained for this object,  which is more consistent
with the 857 GHz value.

\subsection{ERCSC-545 GHz and H-ATLAS $500 \mu m$ flux densities} \label{sec:500vs545}

The comparison between the \emph{Planck} 545 GHz and the SPIRE 500 $\mu$m flux densities is less straightforward. The central frequency of the 500 $\mu$m channel  is $\sim 600$ GHz, significantly higher than that of the nearest \emph{Planck} channel (545 GHz). A colour correction is thus necessary. From the mean SED of IRAS PSC$z$ galaxies determined by \cite{SerjeantHarrison2005}, we find $\langle S_{600}/S_{545}\rangle \simeq 1.35$. For the comparison with \emph{Planck} we scale down the H-ATLAS flux densities by this factor, except for the rich clump of low-flux galaxies, for which we adopted a correction factor of 1.1 for the strongly lensed galaxy at $z=3.259$ and of 1.3 for the surrounding H-ATLAS sources, assumed to be at $z \leq 1$. 
These factors have been calculated using the 
SED of \object{H-ATLAS J142413.9+022304}, a strongly lensed sub-mm galaxy at $z\approx 4.24$ \citep{cox11}, for the strongly lensed galaxy and the SED of \object{SMM J2135-0102} \citep[`The Cosmic Eyelash';][]{Ivison,Swinbank}, that \cite{Lapi2011} found to work well for many high-$z$ H-ATLAS galaxies.
We will denote the colour corrected flux by the symbol $S^*$.

\begin{figure}
  \resizebox{\hsize}{!}{\includegraphics{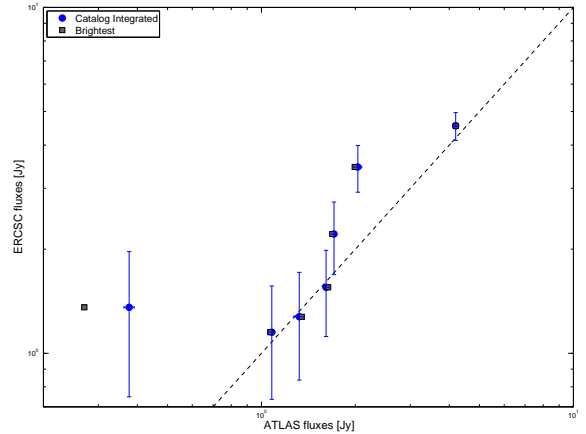}}
  \caption{ERCSC flux densities at 545 GHz compared with the colour-corrected 500 $\mu$m flux densities of the brightest H-ATLAS sources inside the \emph{Planck} beam (squares). The dashed line indicates the $S_{\mathrm{ATLAS}}=S_{\mathrm{ERCSC}}$ identity. The filled circles show the colour-corrected summed flux densities, weighted with a Gaussian beam centered on the ERCSC position and with $\hbox{FWHM}=4.47$ arcmin \citep{ERCSC}. Only the ERCSC sources with at least one \emph{Herschel} counterpart with (non colour-corrected) flux density $S_{\mathrm{\tiny ATLAS}}>0.25$ Jy are shown on this plot. The outlier with ERCSC flux density $\sim 3.5$ Jy is the edge-on spiral galaxy NGC5746. The outlier with ERCSC flux density $\sim 1.5$ Jy is the G12H29 source to be discussed in \S\ 5.}
  \label{fig_flux_comparison_500vs545}
\end{figure}

Figure~\ref{fig_flux_comparison_500vs545} 
compares the \emph{Planck} 545 GHz
flux density with that of the brightest source inside the {\it Planck} beam
(filled squares) and with that obtained summing the flux densities of H-ATLAS sources within the \emph{Planck} beam, corrected for the effect of the beam response ($\hbox{FWHM}=4.47$ arcmin) function and the colour correction (filled circles). Only the \emph{Planck} ERCSC 545 GHz sources that have at least one H-ATLAS counterpart with (non colour corrected) $S_{\rm {\tiny ATLAS}}\gsim 250\,$mJy are shown in the plot. Again NGC5746 and the clump around \object{G270.59+58.52} stand out for their high ERCSC to H-ATLAS flux density ratios. Leaving these aside, we find, after the colour correction, $\langle S^*_{\rm {\tiny ATLAS}}-S_{\rm {\tiny ERCSC}}\rangle = -0.16$ Jy, with a dispersion of $\simeq 0.25\,$Jy for sources with $S^*_{\rm {\tiny ATLAS}}>1\,$Jy. The corresponding relative difference is $\langle \epsilon \rangle = -7.5\%$ with dispersion $\sigma_{\epsilon}=13.5\%$. This result, however, has been obtained from a very small sample of five sources and cannot be considered statistically meaningful.

\section{Contamination by Galactic cirrus} \label{sec:cirrus}

The GAMA-09 field is more contaminated by Galactic emission than the other two GAMA fields \citep{bracco}. None of the 15 ERCSC 857 GHz sources, in this field with $S_{\rm {\tiny ERCSC}}$ in the range 1.2--3.5 Jy, have a plausible \emph{Herschel} counterpart. The summed flux densities of the faint \emph{Herschel}   sources within the \emph{Planck} beam fall well short of accounting for the ERCSC flux densities.
A visual inspection of Figure~\ref{fig:poststamps_GAMA-09} clearly reinforces the idea that the ERCSC sources in this region of the sky are not associated with bright \emph{Herschel} galaxies. 
 All but two of the ERCSC sources in the GAMA-09 have a cirrus flag $\ge 0.125$ and all but two (different from the previous two objects) are labeled as extended in the ERCSC. It is thus likely that most of the flux density within the \emph{Planck} beam comes from Galactic cirrus. The situation is much better in the GAMA-12 and GAMA-15 fields. In the former, only one (out of five) 857 GHz ERCSC source does not have an H-ATLAS counterpart with $S_{\mathrm{\tiny ATLAS}}>250$ mJy. Somewhat surprisingly, this source is not labeled as extended in the ERCSC and has a relatively low cirrus flag (0.0625). One of the two ERCSC 545 GHz sources in the same field behaves in a similar way: no H-ATLAS counterpart with $S_{\rm {\tiny ATLAS}}>250$ mJy, not labeled as extended, low cirrus flag (0.03125). Fortunately, all the eight 857 GHz and the five 545 GHz ERCSC sources in the GAMA-15 field do have a consistent H-ATLAS counterpart.

Although the statistics are poor, these findings may indicate that all ERCSC sources with a cirrus flag $\ge 0.125$ are cirrus dominated, even if they are not labeled as extended, as is the case for two 857 GHz and three 545 GHz GAMA09 sources. Of the three 857 GHz sources labeled as extended but with a cirrus flag $< 0.125$, two are probably cirrus dominated, while the third is the composite  high-$z$ lensed galaxy plus low-$z$ clump (see \S\,\ref{sec:proto}). All 545 GHz sources labeled as extended are probably cirrus dominated, even if the cirrus flag is $< 0.125$.

We can then tentatively conclude that a cirrus flag $\ge 0.125$ {\it or} the `extended' label are good indicators of cirrus dominance, although their presence (or absence) does not always guarantee that a source is (or is not) dominated by cirrus.

%\captionsetup[subfigure]{font=large,
%labelformat=parens,labelsep=space,
%listofformat=subparens}

%\begin{figure}
%  \resizebox{\hsize}{!}{\includegraphics{G12v2.30.eps}}
%  \caption{H-ATLAS 250 $\mu$m image, centered on \object{HATLAS J114637.9-001132}.}
%  \label{fig:zoom}
%\end{figure}

%\begin{figure}
%  \resizebox{\hsize}{!}{\includegraphics{G12v2.30_zoom.eps}}
%  \caption{SDSS sources within the H-ATLAS $500\,\mu$m beam centered on \object{HATLAS J114637.9-001132} (squares). The diamonds are other SDSS sources in the area of $62.94'' \times 39.45''$. A dense clump of foreground galaxies is visible at the center. The two central galaxies have an estimated photometric redshift $z\simeq 1$ \citep{fu}.}
%  \label{fig:SDSS}
%\end{figure}

%\begin{figure}
%  \resizebox{\hsize}{!}{\includegraphics{g12h29-co32_spec2.eps}}
%  \caption{CO spectrum of \object{HATLAS J114637.9-001132} from IRAM 30m showing redshift measurement using CO 3-2 and yielding a redshift of 3.259}
%  \label{fig:zspec}
%\end{figure}

\section{The strongly lensed \textit{Herschel}/\textit{Planck} source H-ATLAS J114637.9-001132 at $z=3.26$} \label{sec:proto}

\begin{figure*}
  \resizebox{\hsize}{!}{
  \centering
  \subfloat{\includegraphics[scale=1]{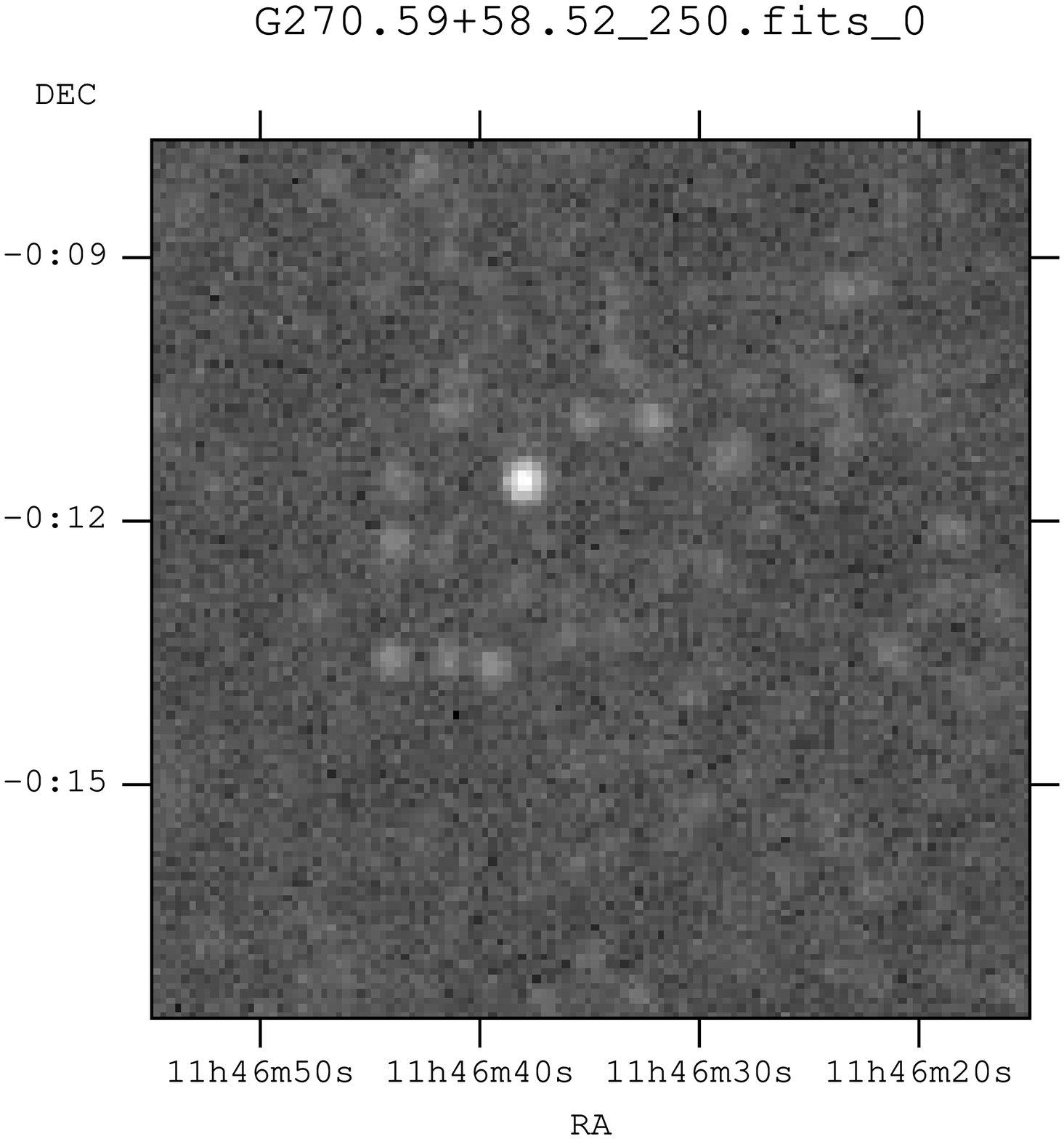}}
  \subfloat{\includegraphics[scale=1]{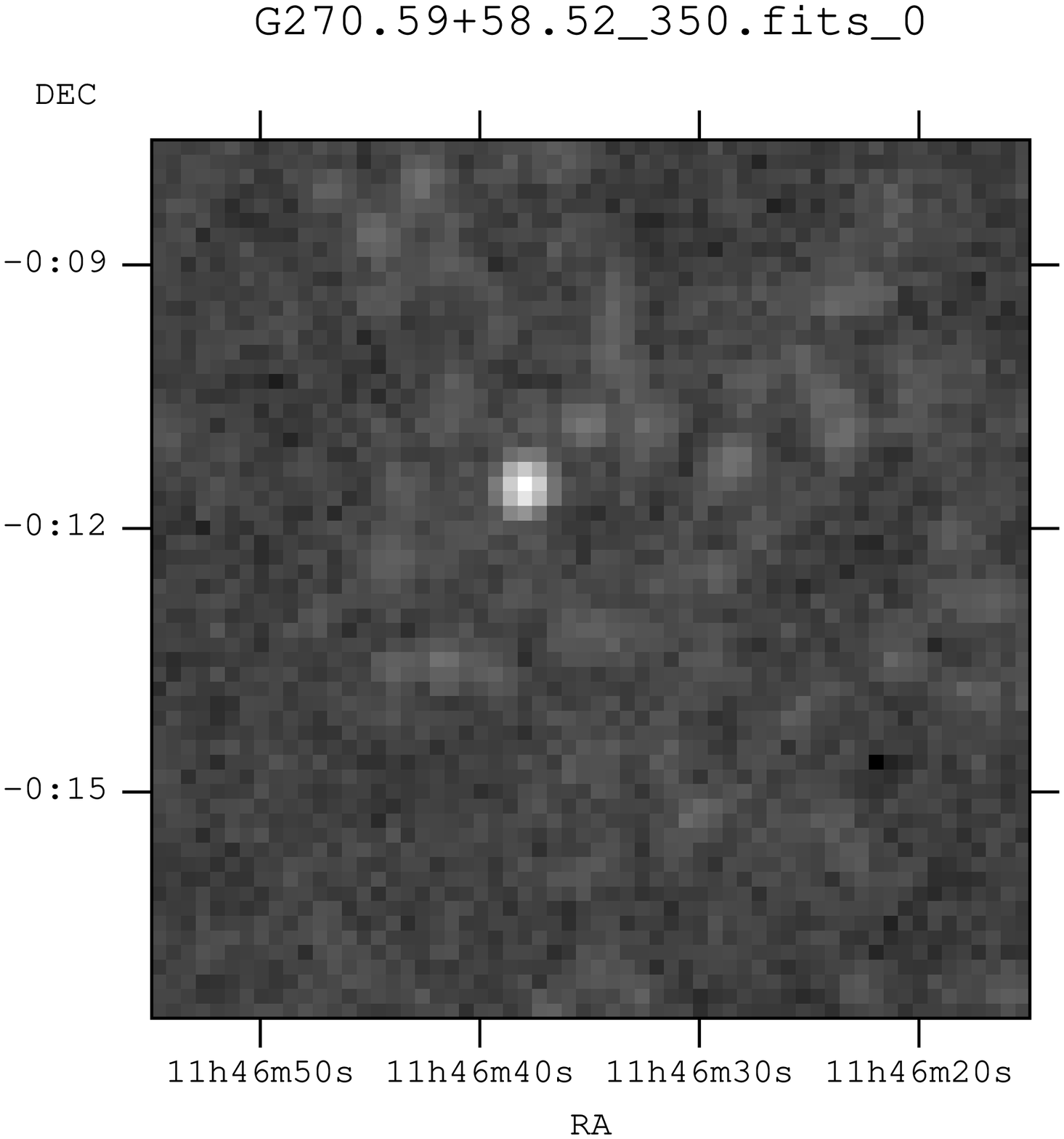}}
  \subfloat{\includegraphics[scale=1]{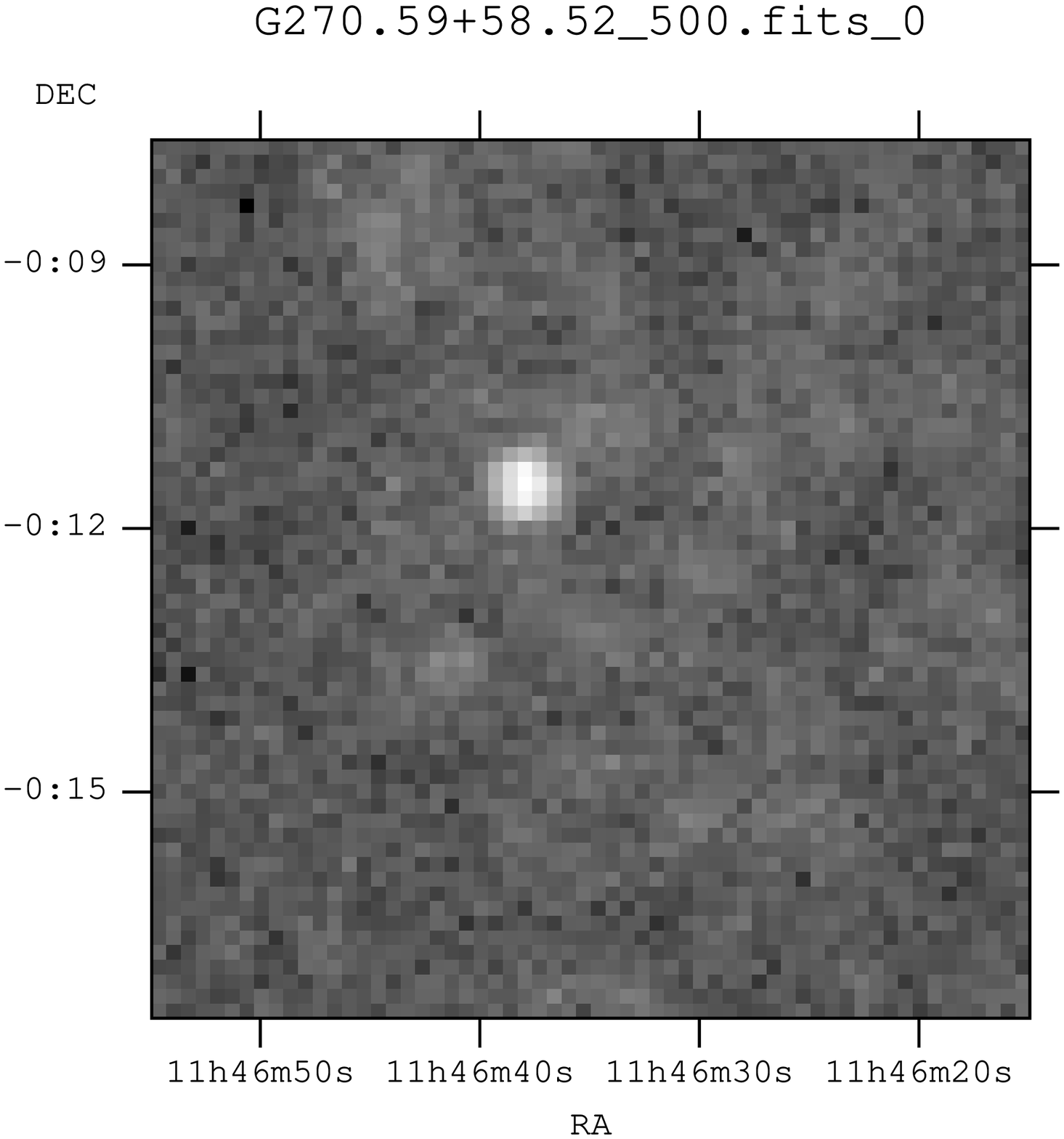}}  }
%  \resizebox{\hsize}{!}{\includegraphics{comparison_500_545.eps}}
  \caption{The clump around G12H29 as seen by \emph{Herschel} at 250 (left), 350 (center) and 500 $\mu$m (right). The field of view is 600 arcsec wide and is centered at the position of the 857 GHz ERCSC source \object{PLCKERC857 G270.59+58.52}.}
  \label{fig:proto}
\end{figure*}

\begin{figure}
 
  \centering
  \includegraphics[width=0.8\columnwidth]{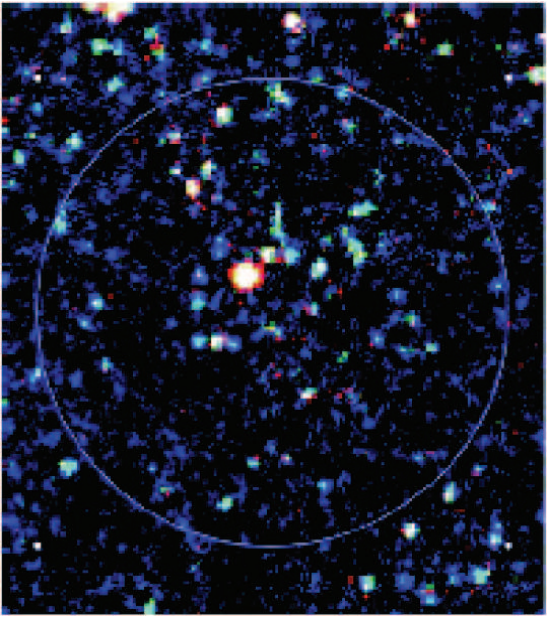}
  \caption{\textit{Herschel} three colour image of the region around \object{HATLAS J114637.9-001132} (alias \object{G12H29}). Colours are: blue -- 250 $\mu$m; green -- 350 $\mu$m; red -- 500 $\mu$m. The colour scale has been chosen so that objects that are red/white are likely to lie at $z\sim3$. The brightest such object in this image is the $z=3.259$ likely lensed galaxy \object{HATLAS J114637.9-001132}. The circle is centered on the \textit{Planck} source and is the size of the \textit{Planck} 857 GHz beam.}
\label{fig:protocol}
\end{figure}

\begin{figure*}[!]
  \resizebox{\hsize}{!}{
%  \centering
  \includegraphics[scale=1,angle=90]{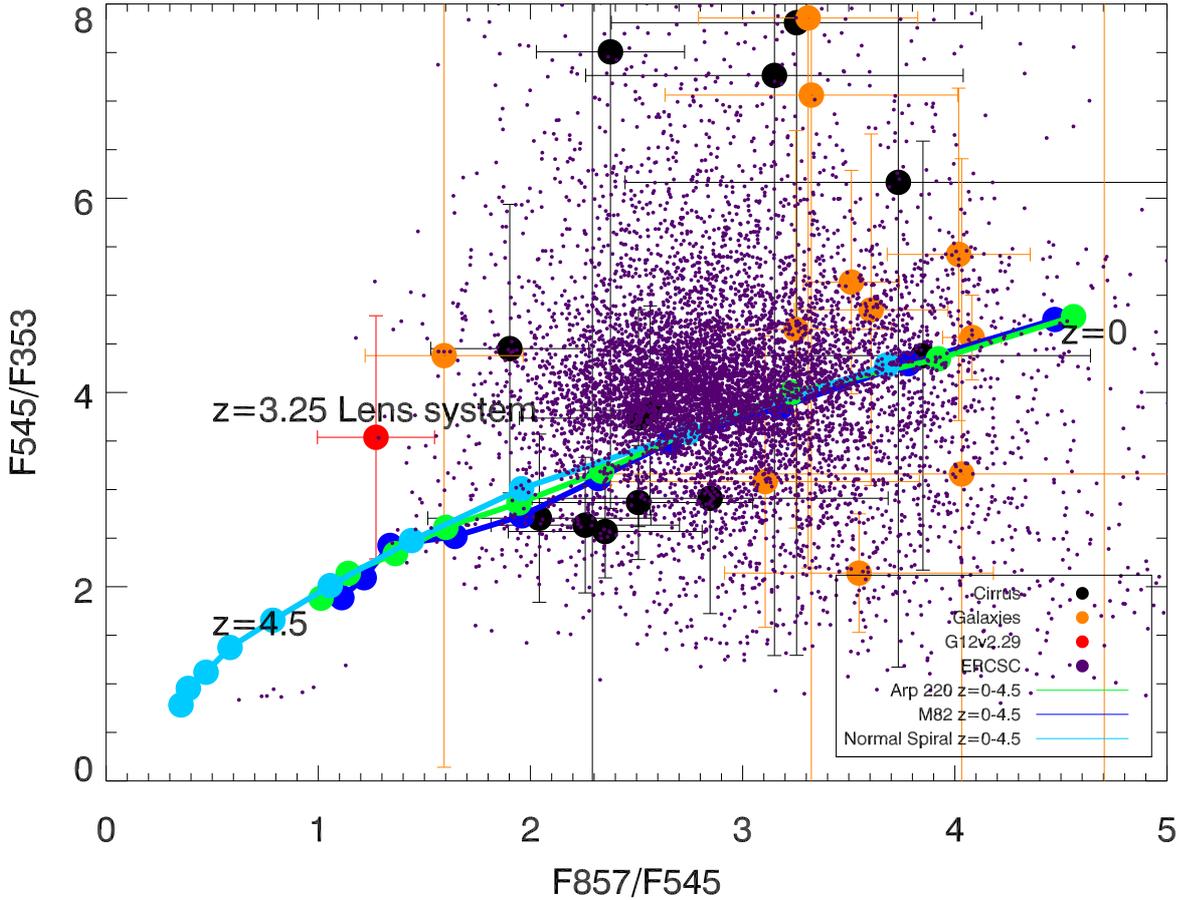}}
%  \resizebox{\hsize}{!}{\includegraphics{comparison_500_545_new.eps}}
  \caption{\textit{Planck} colours of detected objects in the H-ATLAS Phase 1 regions compared to model SED colour tracks for two star-forming and one normal spiral galaxy template. The template SEDs go from $z=0$ to $z=4.5$, with the dots along the lines spaced by $\Delta z=0.5$. The large dots show the ERCSC sources within the ATLAS fields; the 1 sigma error bars for these sources are also included in the plot. A colour code has been assigned  based on our classification using the \emph{Herschel} images: orange dots correspond to sources we classify as galaxies whereas black dots are classified as Galactic cirrus. H-ATLAS J114637.9-001132 (alias G12H29) is shown in red. For comparison, all ERCSC sources with $|b|>20^{\circ}$ and with $3\sigma$ detections or better are also shown as purple points. Seven of the eight isolated purple points in the lower left corner of the plot correspond to blazars identified in the ERCSC \citep{blasas}.}
\label{fig:protocol2}
\end{figure*}

The ERCSC object \object{PLCKERC857 G270.59+58.52}, shown as observed by \textit{Herschel} in panel (c) of Figure~\ref{fig:poststamps_GAMA-12} and in Figure~\ref{fig:proto}, has some unusual characteristics that make it interesting and worthy of further study. Eleven out of the 12 matches between ERCSC sources found at 857 GHz and H-ATLAS sources with $S_{\rm {\tiny ATLAS}}> 250\,$mJy are associated with nearby quiescent galaxies detected in the optical. One source, however, has no bright optical counterpart. Instead, it is associated with a clump of sources with low \textit{Herschel} flux densities (as can be better appreciated in Fig.~\ref{fig:protocol}) grouped around a bright \textit{Herschel} source, \object{HATLAS J114637.9-001132}, alias G12H29, whose flux peaks at 350 $\mu$m. Its \emph{Planck} colour is also unusual, as can be seen in Fig.~\ref{fig:protocol2} where G12H29 is indicated by a red dot. This source was already a target for spectroscopic sub-mm follow up of candidate lensed galaxies \citep[see e.g.][]{negrello10}. A CO spectroscopic redshift of $z=3.259$ has been obtained for this source (Harris et al. 2012, in preparation, Van der Werf et al. 2012, in preparation).
Recent 
LABOCA data  does suggest the presence of other sources in the same clump with SPIRE-to-870 flux ratios that match those of the $z=3.26$ source, providing indirect  evidence for other sources associated with it at the same redshift (Clements et al. in preparation)

The SDSS DR7 (\cite{Abazajian2009}) shows a dense clump just on top of G12H29. \cite{fu} have studied G12H29 and the SDSS clump in detail and determined
the lensing nature of \object{HATLAS J114637.9-001132} during a Keck laser guide star adaptive optics imaging ($J$ and $K_{\rm s}$-band) program of bright Herschel $500 \mu$m sources from H-ATLAS. 
The $K_{\rm s}$-band image shows complex filamentary structures that do not appear at $J$-band. Observations with the SubMillimeter Array (SMA) 
reveal two $880$ $\mu$m sources with flux densities  of 31 mJy and 27 mJy, separated by $3^{\prime \prime}$. In their work, \cite{fu} show that the observations can be nicely explained by a lens model in which the lens is a rather complex system located at $z\sim 1$. 
The photometric redshifts of the optical sources around the lensed object have been determined using SDSS $ugriz$ + UKIDSS $YJHK$ photometry using the publicly available photo-$z$ code EAZY \citep{EAZY}. The Keck $K$-band data were not used for the photo-$z$ calculation. For the two central lensing galaxies, their blended photometry indicates a photo-$z$ of 1.076 (the 68\% confidence interval is 0.982 to 1.305), which is substantially higher than the SDSS DR8 photo-$z$ (0.71). For a more detailed discussion on the lens, the reader is referred to \cite{fu}.

%The total IR luminosity of the candidate lensed galaxy is $L_{\rm IR}\simeq 1.3\times 10^{14}/\mu\,L_\odot$, $\mu$ being the gravitational amplification, if we adopt the SED of \object{H-ATLAS J142413.9+022304} (alias G15.141), a strongly lensed sub-mm galaxy at $z\approx 4.24$ \citep{cox11} that gives a photometric redshift $z_{phot} = 3.17 \pm 0.05$, quite close to the spectroscopic one $z_{spec}=3.259$. If we  use the SED of \object{SMM J2135-0102} \citep[`The Cosmic Eyelash';][]{Ivison,Swinbank}, that \cite{Lapi2011} found to work well for many high-$z$ H-ATLAS galaxies, we find $L_{\mathrm{IR}}\simeq 0.98\times 10^{14}/\mu\,L_\odot$. Comparing the measured CO(1-0) luminosity with that expected from the empirical relationship between line-width and amplification-corrected luminosity for $z\simeq 2$ ultra-luminous infrared galaxies, Harris et al. (2011)  estimate $\mu=6$. Using the relationships between IR luminosity and star-formation rate (SFR) obtained by \cite{Lapi2011} we get  SFRs in the range 1600--$2800(6/\mu)\,M_\odot/$ yr. The corresponding halo mass given by eq.~(9) of \cite{Lapi2011} is $M_H\simeq 1.5\hbox{--} 2.6\times 10^{13}\,M_\odot$. The \cite{ShethTormen1999} halo mass function yields  a very low comoving density of halos that massive ($\log[N(> M_H)/\hbox{Mpc}^3]\simeq -6.27$--$-7.05$) suggesting that the gravitational amplification may have been under-estimated. \textcolor{blue}{AMPLIFICATION ESTIMATE BY HAI.}

Apart from the SDSS clump already described, there is an unusually rich condensation of other low flux H-ATLAS objects within the \emph{Planck} beam centered on G12H29. 
Several of these sources have red \textit{Herschel} colours. We have determined photometric redshifts for these objects using the \object{SMM J2135-0102} SED, that was shown to work quite well for $z\ge 1$ \citep[][Gonz\'alez-Nuevo et al. 2011, in preparation]{Lapi2011}. We find that four out of the six H-ATLAS sources that are located inside a circle of radius $\sigma_{857}=4.23/2\sqrt{2\log 2}$ arcmin \citep{ERCSC} centered on G12H29 have $0.75 < z_{\rm phot} < 1.25$. We have checked that other photo-$z$ methods (eg. Clements et al. b) in preparation) produce answers that are broadly consistent with those from the \cite{Lapi2011} method. However, these photo-$z$ are calculated using only the three SPIRE bands and are therefore uncertain. For those H-ATLAS objects that can be matched to SDSS galaxies, we have recalculated the photo-$z$ using the optical data. Our results seem to indicate that most of these clump sources have lower redshifts $\langle z \rangle < 0.3$ than estimated using SPIRE data alone.

In order to test if the over-density we observe around \object{HATLAS J114637.9-001132} is statistically significant, we have chosen randomly 1000 \emph{Herschel} sources and counted the number of galaxies in circles with the 
same radius $\sigma_{857}$. The median number of neighbours is 1, with a standard deviation of 1.2. None of the sources in the control sample has a number of neighbours equal to or larger than  \object{HATLAS J114637.9-001132}.

Following a bootstrap-type argument and realizing that we can divide the survey area into 30257 independent cells of size $\sigma_{857}$ around H-ATLAS sources (ordered downwards in flux density at 350 microns), we have counted the number of H-ATLAS sources inside each one of these cells. We find that the fraction of cells as populated (or more) as the G12H29 clump is $9.915 \times 10^{-4}$ (30 cells). 
Using a
Bayesian approach and the binomial distribution \citep{wall2003}, we find that the 99.99\% confidence interval
for that fraction is   $[0.4422\times 10^{-4} , 1.8736\times 10^{-3}]$ (narrowest
interval that includes the mode and encompasses 99.99\% of the
probability, S. Andreon, private communication). 
We conclude that the over-density observed around G12H29 is statistically significant, but given our limited knowledge about the redshift distribution of the clump objects it is not clear whether the clump is a real association of objects at the same redshift or a random alignment of galaxies at very different distances. Our current data seem to favour this last interpretation (a low redshift clump at $z<0.3$ plus the lens SDSS clump at $z\sim 1$ plus the lensed galaxy at $z=3.259$), but the large uncertainties of photo-$z$ estimates make it impossible to rule out other possibilities at this point.

The large \textit{Planck} beam means that there is considerable potential for source confusion to affect the colours. The H-ATLAS survey has detected at $\ge 5\sigma$ in the \textit{Herschel} 350 $\mu$m band (equivalent to the \textit{Planck} 857GHz channel), sources with a total flux density of $S_{\rm {\tiny ATLAS}} = (0.52 \pm 0.01)\,$Jy (taking into account the effect of a Gaussian beam with $\hbox{FWHM}=4.23\,$arcmin centered on the \textit{Planck} position) to be compared with the ERCSC flux density $S_{\rm {\tiny ERCSC}} = 2.15\pm0.83\,$Jy. 
Again allowing for the effect of the \textit{Planck} beam (with $\hbox{FWHM}=4.47\,$arcmin in this case), the summed contributions of $\ge 5\sigma$ H-ATLAS sources at $500\mu$m is $S_{\rm {\tiny ATLAS}} = (0.44 \pm 0.02)\,$Jy. Applying a colour correction of a factor of 1.1 for G12H29 and 1.3 for the other sources (assumed to be at redshift $\leq 1$) we get $S^*_{\rm {\tiny ATLAS}} = (0.38 \pm 0.01)\,$Jy, compared with $S_{\rm {\tiny ERCSC}} = (1.36\pm0.61)\,$Jy. This suggests that \emph{Planck} measurements are boosted by a positive background fluctuation, which may also account for the fact that only one (out of 9) H-ATLAS sources with flux densities at 350 $\mu$m $\simeq 300$ mJy and $z \geq 1$, over the entire H-ATLAS Phase 1 area, is associated with a \emph{Planck} detection. In fact, background fluctuations of the order of $\geq 2$ sigma are needed in order to make those sources detectable by \emph{Planck}. The probability of such an event is so extremely low (just a few per cent for Gaussian fluctuations) that even a single occurrence can be considered a stroke of luck. Since the fluctuations are dominated by confusion and have a strong super-Gaussian tail, the frequency of these fluctuations is substantially higher than expected from Gaussian statistics. However, the probability that this fluctuation happens to be by chance on top of a strongly lensed galaxy is tiny (but hard to quantify without knowing the statistics of fluctuations). Therefore the most likely scenario is the one in which H-ATLAS J114637.9-001132 is associated with a clump of sources, most of which fall below the H-ATLAS detection limit but their total integrated flux is seen as a positive fluctuation by \emph{Planck}, due to its relatively large beam.

Figure~\ref{fig:protocol2} shows the distribution of   $F_{545}/F_{353}$ vs. $F_{857}/F_{545}$ colours for the ERCSC sources in the H-ATLAS fields (black, orange and big red dots) and in the rest of the sample (small purple dots), compared to model SED colour tracks for two star-forming and one
normal spiral galaxy templates going from $z = 0$ to $z = 4.5$ with a $z=0.5$ interval. The black dots correspond to the sources we have classified as Galactic cirrus, whereas orange dots are thought to be truly extragalactic. The red dot corresponds to G12H29. Its isolated position in the diagram suggests that there will be very few other objects like this in the ERCSC\footnote{Another interesting feature of this diagram, somewhat apart from the focus of this paper, are the eight isolated purple dots that appear in the lower left part of the plot: seven of them correspond to blazars identified in the ERCSC \citep{blasas}.}. 

\section{Conclusions} \label{sec:conclusions}

%\textcolor{blue}{Here be conclusions!}
A cross-correlation of the \emph{Planck} Early Release Compact Source Catalog (ERCSC) with the catalogue of \emph{Herschel}-ATLAS sources detected in the Phase 1 fields, covering $134.55\,\hbox{deg}^2$, has highlighted several issues that need to be dealt with to correctly interpret the data from the \emph{Planck} sub-mm surveys.

\begin{itemize}

\item Contamination by diffuse Galactic emission is a serious problem, as demonstrated by the fact that even in a region of moderate Galactic emission (GAMA09) all the 857 GHz ERCSC sources seem to be associated with cirrus. Therefore, to estimate e.g. the number counts of extragalactic sources, it is crucial to carefully select regions of low Galactic emission. A cirrus flag $\ge 0.125$ and the `extended' flag, as defined in \cite{ERCSC}, are remarkably effective in picking up probable cirrus, but are not 100\% reliable.

\item We find a good, essentially linear, correlation between ERCSC flux densities at 857 GHz and SPIRE flux densities at 350 $\mu$m above $S_{\rm {\tiny ERCSC}} \simeq 1.5\,$Jy. We also find a good correlation between ERCSC flux densities at 545 GHz for sources $S_{\rm {\tiny ERCSC}} \geq 1\,$Jy
and SPIRE flux densities at 500 $\mu$m, after a colour correction has been applied to SPIRE flux densities in order to  take into account the different central wavelenghts of the bands.   Excluding the large edge-on disk galaxy NGC5746, whose H-ATLAS flux density is different form the ERCSC values probably due to resolution and background subtraction systematic effects,  we find $\langle S_{\rm {\tiny ATLAS}}-S_{\rm {\tiny ERCSC}}\rangle\simeq 0.1\,$Jy at 857 GHz and $\langle S^*_{\rm {\tiny ATLAS}}-S_{\rm {\tiny ERCSC}}\rangle = -0.16$ Jy at 545 GHz. ERCSC flux densities are affected by flux boosting  and have $\geq 30\%$ uncertainties below $\sim 1.3$ Jy.   
The relative difference between \emph{Herschel} and \emph{Planck} flux densities is compatible with these error levels and the  calibration uncertainties of both experiments. The mean difference between ERCSC and H-ATLAS positions for 857 GHz sources brighter than 1.5 Jy is 0.42 arcmin, with a dispersion of 0.17 arcmin, confirming the accuracy of ERCSC positions.

\item Apart from the contamination from Galactic thermal dust emission, the \emph{Planck} sub-mm surveys are limited by confusion due to faint sources within the beam, as expected \citep[e.g.][]{Negrello2004,fconde08}. An important contribution to confusion fluctuations is clustering \citep{PlanckCIB}. Occasionally the confusion fluctuations may be dominated by a single proto-cluster of star-forming galaxies \citep{negrello05}. Even less frequently the confusion fluctuations may be due to apparent clustering due to a random alignment of galaxies at different redshifts. 
 We have presented evidence suggesting that at least one object with an anomalous contamination from confusion fluctuations 
  has been detected within the H-ATLAS Phase 1 fields. This source is a mixture of a strongly lensed galaxy at $z=3.26$ surrounded by a statistically significant overdensity of faint galaxies detected by SPIRE.  This unusual overdensity of faint galaxies plus an excess of confusion fluctuations at the same position has made it possible for the ERCSC to detect a high redshift lensed galaxy that would otherwise be below the \emph{Planck} detection limit.  
  The available information is insufficient to reliably estimate the redshift of the galaxies in this clump, although there are some indications that at least the lensing galaxies are at $z\sim 1$. Upcoming PACS photometry and near-IR follow-up of the galaxies in this clump will allow us to better constrain the photometric redshifts of its galaxies.   If the rest of  the galaxies of the clump could be shown to be at the same redshift, then it would represent an example showing  the power of combining \emph{Planck} and \emph{Herschel} data. Such a combination may open a new window for the study of cluster evolution, since the main searches carried out so far at similar redshifts have selection functions that are different to that of sub-mm surveys. X-ray and SZ surveys preferentially find massive and evolved structures, dominated by passive early type galaxies. Optical/nearIR cluster finding algorithms, depending on what detection technique is used, can be biased to red evolved galaxies (e.g. Red Sequence fitting) but this is not always the case \citep{vanBreukelen}. 
The sub-mm selection could find clusters with a high level of star-formation activity, thus shedding light on this poorly known phase of their evolution \citep{micha}.

\item  Although our statistics are too poor to come to definite conclusions, simple source blending seems to be a less frequent problem: in only one case has an ERCSC 857 GHz source been resolved by \emph{Herschel} into two similarly bright objects, and in general, the contribution of lower luminosity H-ATLAS sources to the flux density measured by \emph{Planck} was minor (see Fig.~\ref{fig_flux_comparison_350vs857}).

\end{itemize}

In conclusion, we find that the higher sensitivity and higher angular resolution H-ATLAS maps provide key information for the interpretation of candidate sources extracted from \emph{Planck} sub-mm maps. The  Phase 1 survey considered in this paper represents  $\sim 1/4$ of the full H-ATLAS ($550\,\hbox{deg}^2$). Therefore, the results presented here will be substantially improved upon when the H-ATLAS survey is completed.

Of special interest is the possibility that \emph{Planck} may be able to sample the tail of the distribution of high-$z$ over-densities, providing unique information about both the the early evolution of large-scale structure and galaxy formation and evolution in high density environments.

\begin{acknowledgements}
The \textit{Herschel}-ATLAS is a project with \textit{Herschel}, which is an ESA space observatory with science instruments provided by European-led Principal Investigator consortia and with important participation from NASA. The \textit{H-ATLAS} website is http://www.h-atlas.org/. DH and MLC acknowledge partial financial support from the Spanish
Ministerio de Ciencia e Innovaci\'on project AYA2010-21766-C03-01 and the Consolider Ingenio-2010
Programme project CSD2010-00064. DH also acknowledges the Spanish Ministerio de Educaci\'on for a Jos\'e Castillejo' mobility grant with reference JC2010-0096 and the Astronomy Department at the Cavendish Laboratory for their hospitality during the elaboration of this paper.  The Italian group has been supported in part by ASI/INAF agreement n. I/009/10/0 and by INAF through the PRIN 2009 ``New light on the early Universe with sub-mm spectroscopy''. FJC acknowledges partial financial support from the Spanish Ministerio de Ciencia e Innovaci\'on project AYA2010-21490-C02-01.
\end{acknowledgements}

\bibliographystyle{aa}
\bibliography{bibfile}

\clearpage
\onecolumn

\begin{landscape}
\begin{table}
\centering
\tiny
\begin{tabular}{ l l l l l r l l l l l r l }
\hline
  \multicolumn{1}{ c }{ERCSC NAME} &
  \multicolumn{1}{c }{RA (deg)} &
  \multicolumn{1}{c }{DEC (deg)} &
  \multicolumn{1}{c }{$S_{857\rm GHz}$ (Jy)} &
  \multicolumn{1}{c }{$\Delta S_{857\rm GHz}$ (Jy)} &
  \multicolumn{1}{c }{EXT} &
  \multicolumn{1}{c }{Cirrus} &
  \multicolumn{1}{c }{H-ATLAS ID} &
  \multicolumn{1}{c }{Other ID} &
  \multicolumn{1}{c }{$S_{350\mu{\rm m}}$ (Jy)} &
  \multicolumn{1}{c }{$\Delta S_{350\mu{\rm m}}$ (Jy)} &
  \multicolumn{1}{c }{$z_{\rm spec}$} &
  \multicolumn{1}{c }{D (arcmin)} \\
\hline
 G223.40+22.96 & 127.935 & 1.659 & 10.751 & 1.940 & 1 & 0.219 &                    &            &       &       &       &      \\
 G224.33+24.38 & 129.583 & 1.602 & 9.924 & 1.209 & 1 & 0.188 &                     &           &       &       &       &      \\
 G224.70+24.60 & 129.940 & 1.415 & 6.077 & 0.894 & 1 & 0.156 &                   &            &       &       &       &      \\
 G224.73+23.79 & 129.253 & 0.998 & 8.440 & 1.650 & 1 & 0.156 &                    &            &       &       &       &      \\
 G226.70+24.89 & 131.101 & -0.013 & 6.987 & 1.392 & 1 & 0.125 &                   &             &       &       &       &      \\
 G226.98+26.25 & 132.398 & 0.454 & 6.256 & 1.261 & 1 & 0.109 &                    &            &       &       &       &      \\
 G227.24+24.51 & 131.018 & -0.629 & 7.291 & 0.895 & 1 & 0.125 &                   &             &       &       &       &      \\
 G227.26+24.72 & 131.206 & -0.540 & 3.234 & 0.917 & 1 & 0.125 &                   &             &       &       &       &      \\
 G227.58+24.57 & 131.222 & -0.863 & 8.093 & 1.194 & 1 & 0.125 &                   &             &       &       &       &      \\
 G227.75+30.24 & 136.180 & 1.886 & 13.259 & 2.118 & 1 & 0.078 &                   &             &       &       &       &      \\
 G230.55+31.91 & 138.859 & 0.711 & 1.535 & 0.406 & 0 & 0.125 &                    &            &       &       &       &      \\
 G230.97+32.31 & 139.381 & 0.614 & 0.872 & 1.232 & 1 & 0.125 &                    &            &       &       &       &      \\
 G231.25+32.05 & 139.284 & 0.279 & 3.429 & 0.769 & 1 & 0.125 &                    &            &       &       &       &      \\
 G231.38+32.24 & 139.511 & 0.291 & 2.863 & 2.249 & 1 & 0.125 &                    &            &       &       &       &      \\
 G231.43+32.10 & 139.413 & 0.173 & 4.038 & 0.539 & 0 & 0.125 &                    &            &       &       &       &      \\
 G263.84+57.55 & 173.088 & 0.810 & 1.990 & 0.434 & 0 & 0.047 & J113221.5+004814   &   NGC3720         & 1.2149 & 0.0364 & 0.0198 & 0.405\\
               &         &       &       &       &   &       & J113213.3+004907   &   NGC3719         & 1.0098 & 0.0345 & 0.0195 & 1.998\\
 G266.26+58.99 & 174.940 & 1.323 & 2.056 & 0.667 & 0 & 0.062 &                    &            &       &       &       &      \\
 G270.59+58.52 & 176.646 & -0.211 & 2.145 & 0.824 & 1 & 0.031 & J114637.9-001132  &   G12H29         & 0.3783 & 0.0074 & 3.259 & 1.359\\
 G274.04+60.90 & 179.271 & 1.115 & 1.511 & 0.380 & 0 & 0.109 & J115705.9+010730   &  CGCG 013-010          & 1.2073 & 0.0304 & 0.0395 & 0.628\\
 G277.37+59.21 & 180.100 & -1.104 & 17.578 & 0.441 & 0 & 0.047 & J120023.2-010600 &   NGC4030         & 18.3014 & 0.1060 & 0.0048 & 0.303\\
 G345.11+54.84 & 215.605 & -0.395 & 3.361 & 0.466 & 0 & 0.016 & J142223.4-002313  &   NGC5584          & 3.9080 & 0.0565 & 0.0055 & 0.645\\
 G347.77+56.35 & 215.865 & 1.720 & 1.329 & 0.447 & 0 & 0.031 & J142327.2+014335   &   UGC9215         & 1.6225 & 0.0584 & 0.0046 & 0.412\\
 G350.46+51.85 & 219.962 & -0.716 & 2.235 & 0.596 & 0 & 0.078 & J143949.5-004305  &   NGC5705 	          & 1.4737 & 0.0565 & 0.0059 & 0.381\\
 G351.01+52.11 & 220.048 & -0.298 & 7.824 & 0.635 & 0 & 0.078 & J144011.1-001725  &   NGC5713          & 7.7971 & 0.0739 & 0.0063 & 0.474\\
 G351.22+51.97 & 220.238 & -0.320 & 4.881 & 0.476 & 0 & 0.078 & J144056.2-001906  &   NGC5719          & 5.6896 & 0.1303 & 0.0057 & 0.231\\
 G353.15+54.45 & 219.422 & 2.288 & 6.179 & 0.506 & 0 & 0.062 & J143740.9+021729   &   NGC5690         & 6.4066 & 0.0855 & 0.005847 & 0.239\\
 G354.50+52.84 & 221.113 & 1.676 & 2.942 & 0.511 & 0 & 0.078 & J144424.3+014046   &   NGC5740         & 2.8600 & 0.0474 & 0.0052 & 0.722\\
 G354.96+52.95 & 221.237 & 1.951 & 11.238 & 0.560 & 0 & 0.062 & J144455.9+015719  &   NGC5746          & 7.9449 & 0.1451 & 0.0057 & 0.359\\
\hline
\end{tabular}
\caption{ERCSC sources at 857 GHz and their H-ATLAS counterparts (inside a 4.23 arcmin radius circle around the ERCSC position) at 350 $\mu$m. \emph{Planck} flux densities and their associated errors are taken from the ERCSC GAUFLUX column. Only the H-ATLAS sources with flux density $S_{350\mu{\rm m}}>0.3$ Jy have been included in the table (blank spaces mean that no H-ATLAS source brighter than 0.3 Jy at $350\, \mu{\rm m}$ are associated to the ERCSC detection). Object PLCKERC G263.84+57.55 is a blend of 2 H-ATLAS sources, each with $S_{350\mu{\rm m}}>0.3$ Jy. All the redshifts are spectroscopic. In the case of G270.59+58.52 the redshift refers to the strongly lensed galaxy G12H29. }\label{table2}

\end{table}
\end{landscape}

\clearpage
\onecolumn

\begin{landscape}
\begin{table}
\centering
\tiny

\begin{tabular}{ l c c c c c c l l c c c c }
\hline
  \multicolumn{1}{ c }{NAME} &
  \multicolumn{1}{c }{RA (deg)} &
  \multicolumn{1}{c }{DEC (deg)} &
  \multicolumn{1}{c }{$S_{545\rm GHz}$ (Jy)} &
  \multicolumn{1}{c }{$\Delta S_{545\rm GHz}$} &
  \multicolumn{1}{c }{EXT} &
  \multicolumn{1}{c }{CIRRUS} &
  \multicolumn{1}{c }{H-ATLAS ID} &
   \multicolumn{1}{c }{Other ID} &	
  \multicolumn{1}{c }{$S_{500\mu\rm m}$ (Jy)} &
  \multicolumn{1}{c }{$\Delta S_{500\mu\rm m}$ (Jy)} &
  \multicolumn{1}{c }{$z_{\rm spec}$} &
  \multicolumn{1}{c }{D (arcmin)} \\
\hline
  G224.70+24.61 & 129.951 & 1.422 & 1.543 & 0.517 & 0 & 0.156 &                     &         &      &      &        &     \\
  G226.97+26.25 & 132.394 & 0.458  & 2.185 & 0.647 & 1  & 0.109  &                   &         &      &      &        &     \\
  G227.27+24.71 & 131.198 & -0.550 & 3.718 & 1.536 & 1 & 0.125 &                     &         &      &      &        &     \\
  G227.57+24.56 & 131.215 & -0.863 & 4.234 & 1.194 & 1 & 0.125 &                     &         &      &      &        &     \\
  G227.77+30.23 & 136.181 & 1.869 & 4.554 & 1.168 & 1 & 0.078 &                      &        &      &      &        &     \\
  G230.17+32.05 & 138.809 & 1.055 & 1.191 & 0.463 & 0 & 0.141  &                     &         &      &      &        &     \\
  G231.44+32.10 & 139.420 & 0.174  & 1.410 & 0.389 & 0 & 0.125  &                    &         &      &      &        &     \\
  G270.59+58.54 & 176.648 & -0.199 & 1.358 & 0.610 & 0 & 0.031 & J114637.9-001132    &  G12H29       & 0.298 & 0.008 & 3.259 & 0.746\\
  G277.36+59.21 & 180.094 & -1.097 & 4.546 & 0.419 & 0 & 0.047 & J120023.2-010600    &  NGC4030      & 5.661 & 0.052 & 0.0048 & 0.229\\
  G345.12+54.85 & 215.603 & -0.381 & 1.151 & 0.415 & 0 & 0.016 & J142223.4-002313    &  NGC5584       & 1.441 & 0.033 & 0.0055 & 0.460\\
  G351.01+52.12 & 220.047 & -0.292 & 2.213 & 0.524 & 0 & 0.078 & J144011.1-001725    &  NGC5713       & 2.278 & 0.035 & 0.0063 & 0.141\\
  G351.21+51.96 & 220.242 & -0.327 & 1.275 & 0.439 & 0 & 0.078 & J144056.2-001906    &  NGC5719       & 1.817 & 0.073 & 0.0057 & 0.729\\
  G353.14+54.45 & 219.414 & 2.290 & 1.551 & 0.433 & 0 & 0.062 & J143740.9+021729     &  NGC5690      & 2.210 & 0.046 & 0.0855 & 0.409\\
  G354.97+52.95 & 221.236 & 1.956 & 3.456 & 0.534 & 0 & 0.062 & J144455.9+015719     &  NGC5746      & 2.686 & 0.081 & 0.0058 & 0.204\\
\hline\end{tabular}

\caption{ERCSC sources at 545 GHz and their H-ATLAS 500 $\mu m$ counterparts inside a 4.47 arcmin radius circle around the ERCSC position. \emph{Planck} flux densities and their associated errors are taken from the ERCSC GAUFLUX column. Only the H-ATLAS sources with flux density $S_{500\mu{\rm m}}\gsim 0.3$ Jy have been included in the table (blank spaces mean that no H-ATLAS source brighter than 0.3 Jy at $500\, \mu{\rm m}$ are associated to the ERCSC detection).  In the case of G270.59+58.54 the redshift refers to the strongly lensed galaxy G12H29}\label{table3}

\end{table}
\end{landscape}

%\clearpage
%\onecolumn
%\input{clump_galaxies_table}

\end{document}